\documentclass[aps,twocolumn,showpacs,preprintnumbers,amsmath,amssymb,nofootinbib,superscriptaddress,showkeys]{revtex4-1}

\usepackage{graphicx}
\usepackage{subfigure}
\usepackage{bm,tikz}
\usepackage{slashed} 

\makeatletter

\begin{document}

\title{Generalized quasi, Ioffe-time, and pseudo quark distributions of the pion \\ in the Nambu--Jona-Lasinio model}

\author{Vanamali Shastry}
\email{vshastry@ujk.edu.pl}
\affiliation{Institute of Physics, Jan Kochanowski University, 25-406 Kielce, Poland}

\author{Wojciech Broniowski}
\email{Wojciech.Broniowski@ifj.edu.pl}
\affiliation{H. Niewodnicza\'nski Institute of Nuclear Physics PAN, 31-342 Cracow, Poland}
\affiliation{Institute of Physics, Jan Kochanowski University, 25-406 Kielce, Poland}

\author{Enrique Ruiz Arriola}
\email{earriola@ugr.es}
\affiliation{Departamento de F\'{\i}sica At\'{o}mica, Molecular y Nuclear and Instituto Carlos I de  F{\'\i}sica Te\'orica y Computacional, Universidad de Granada, E-18071 Granada, Spain}

\date{7 September 2022}  
 
\begin{abstract}
We analyze the generalized quasi, Ioffe-time, and pseudo
distributions of the valence quarks in the pion at the quark model
scale. We use the framework of the Nambu-Jona-Lasinio model and
investigate the basic question of how fast the pion has to move to
effectively reach the infinite momentum limit, where the approach
can provide the information on the generalized parton distribution
functions. We consider both the vector distributions and the
transversity distributions, related to the spin densities. With the
developed analytic expressions, we conclude that to effectively
approach the infinite momentum limit in the Ioffe-time distributions 
for values of the Ioffe-time accessible in lattice QCD, one roughly needs the pion
momenta of the order of $\sim 3$~GeV. We explore polynomiality of
the quasi distributions and study the generalized quasi form
factors. The issue of separability of the transverse and longitudinal
dynamics in the model is studied with the help of the generalized
Ioffe-time distributions, with the conclusion that the breaking is not
substantial, unless the momentum transfer $t$ is large. We also
provide an estimate of the range of the Ioffe-time values needed to
obtain the generalized parton distributions with a reasonable
accuracy. Our model results, which are analytic or semi-analytic,
provide a valuable insight into the theoretical formalism and
illustrate the intricate features of the investigated distributions.
\end{abstract}

\maketitle

\section{Introduction \label{sec:intro}}

In recent years a significant progress has been made to directly
access partonic distribution functions (PDFs) of hadrons on the
Euclidean lattices.  The novel techniques involve the
quasi-distributions proposed by
Ji~\cite{Ji:2013dva,Chen:2016utp,Alexandrou:2015rja,Alexandrou:2017dzj,Zhang:2018diq,Izubuchi:2019lyk,Lin:2020ssv,Gao:2020ito,Chen:2020arf,Braun:2020ymy}
amended with the large mass effective theory
(LaMET)~\cite{Ji:2014gla,Chen:2018fwa,Wang:2019tgg,Ji:2020ect}, the
pseudo-distributions~\cite{Radyushkin:2016hsy,Radyushkin:2017cyf,Orginos:2017kos,Monahan:2016bvm,Monahan:2017oof,Radyushkin:2018nbf,Karpie:2018zaz,Joo:2019bzr,Joo:2019jct,Karpie:2019eiq,DelDebbio:2020rgv,Joo:2020spy,Bhat:2020ktg},
the ``good lattice cross sections''
method~\cite{Ma:2014jla,Ma:2017pxb}, or the Compton Feynman-Hellman
approach~\cite{Chambers:2017dov}.  These methods have an ambition to
reach way further than the lattice QCD determinations of the lowest Bjorken
$x$-moments of the hadronic
PDFs~\cite{Martinelli:1987zd,Morelli:1991gb,Best:1997qp,Detmold:2003tm}
and obtain the distributions as functions of the parton momentum
fraction $x$ ($x$ is the momentum fraction of the hadron carried by
the parton). The quasi distribution amplitude (qDA) of the pion on the
lattice has recently been analyzed in~\cite{Gao:2022vyh}. For detailed
reviews of the progress and the challenges, both for the nucleon and
for mesons,
see~\cite{Lin:2017snn,Cichy:2018mum,Monahan:2018euv,Zhao:2018fyu,Amoroso:2022eow}.
One should remark here that apart from the above listed methods, the
valence PDF and the distribution amplitude (DA) of the pion have also
been determined directly within the Hamiltonian transverse lattice
approach which faces the problem in the Minkowski space~\cite{Burkardt:2001mf,Burkardt:2001jg,Dalley:2002nj}.

Model determinations of the quasi parton distributions of the pion have been made in the framework of the Nambu--Jona-Lasinio (NJL) model (for a review of NJL in the context of high-energy processes see~\cite{RuizArriola:2002wr}) in~\cite{Broniowski:2017wbr,Broniowski:2017gfp}, with an extension to the Ioffe-time distributions (ITDs)~\cite{Braun:1994jq,Radyushkin:2017cyf} provided in~\cite{Broniowski:2017zqz}. An analysis within a QCD instanton vacuum model has been carried out in~\cite{Kock:2020frx}. 

For the nucleon, a study in a diquark spectator model was carried out in~\cite{Gamberg:2014zwa}, a factorization ansatz was used in~\cite{Broniowski:2017gfp}, a large-$N_c$ model was explored in~\cite{Son:2019ghf,Son:2022qro}, and the quark quasi Sivers and quasi Boer-Mulders functions were considered in~\cite{Tan:2022kgj}.

On the experimental side, the current and indirect knowledge of
the pion's valence PDF originates from scattering of secondary pion
beams on nuclear targets at the CERN NA3 experiment~\cite{NA3:1983ejh}
and the Fermilab E615 experiment~\cite{Conway:1989fs}, as well as from
the electroproduction HERA data~\cite{ZEUS:2002gig,H1:2010hym}. A
recent global analysis has been presented in~\cite{Barry:2021osv}.
The future COMPASS++/AMBER facility at CERN, with direct pion beams,
will provide separate information on the valence and sea pion PDFs, as
well as on the gluon distribution, using the J/$\psi$ and $\psi'$
production~\cite{Adams:2018pwt}. Hence, there emerges a renewed demand for theoretical and model predictions.

An evaluation of the pion PDFs in NJL, keeping track of relativity, gauge invariance, a proper support in the Bjorken variable $x$, induced normalization constraints, and supplemented with the QCD evolution, was first made
in~\cite{Davidson:1994uv,Davidson:2001cc} thanks to a scrupulous
  implementation of a suitable regularization method. Analyses in
non-local chiral quark models were provided
in~\cite{Dorokhov:2000gu,Anikin:2000rq,Noguera:2005cc}.  For the
results of the Dyson-Schwinger approach with the rainbow ladder
truncation, see~\cite{Nguyen:2011jy,Chang:2014lva}.

Extensions of PDFs to non-forward processes~\cite{Bartels:1981jh,Geyer:1985vw,Dittes:1988xz} are provided by the generalized parton distribution functions (GPDs)~\cite{Ji:1996nm,Radyushkin:1997ki}, where the momenta of the initial and final hadron can be different. 
Naturally, this enriches the insight into the hadron structure, leading to partonic tomography revealing the spatial distribution of partons in the plane transverse to the direction of motion of the hadron~\cite{Burkardt:2000za,Burkardt:2002hr,Burkardt:2015qoa,Berthou:2015oaw}. For extensive reviews of GPDs see, e.g.~\cite{Ji:1998pc,Radyushkin:2000uy,Goeke:2001tz,Bakulev:2000eb,Diehl:2003ny,Ji:2004gf,Tiburzi:2004ye,Belitsky:2005qn,Boffi:2007yc,Feldmann:2007zz,Boffi:2007yc,Boer:2011fh,Guidal:2013rya}, where also the significance of GPDs to physical processes, such as deeply virtual Compton scattering (DVCS) or the hard meson production (HMP) is reported. 

The pion GPDs in a non-local chiral quark model were presented in~\cite{Praszalowicz:2002ct}, whereas the NJL results can be found in~\cite{Theussl:2002xp,Broniowski:2007si,Courtoy:2010qn,Dorokhov:2011ew,Zhang:2021shm}. In particular, analytic expressions at the quark model scale were presented in~\cite{Broniowski:2007si}. For an evaluation of the GPDs of light mesons in the light-front Hamiltonian approach see~\cite{Adhikari:2021jrh}. The $\rho$ meson GPD in the NJL model was shown in~\cite{Zhang:2022zim}.
Transverse lattice calculations were given in~\cite{Dalley:2003sz}, with the results reproduced by an NJL calculation~\cite{Broniowski:2003rp}.

Recently, the accessibility of the pion GPD through the Sullivan
process~\cite{PhysRevD.5.1732} has been brought up in the context of
the future Electron-Ion Collider
experiments~\cite{Aguilar:2019teb,Chavez:2021llq}.

A path to obtain GPDs on the Euclidean lattice from the quasi GPDs
(qGPDs) has been proposed in~\cite{Ji:2015qla,Liu:2019urm}, and an
alternative approach based on the pseudo GPDs (pGPDs) has been
advocated in~\cite{Radyushkin:2019owq}.  In a quark spectator model,
qGPDs were addressed
in~\cite{Bhattacharya:2018zxi,Ma:2019agv,Bhattacharya:2019cme},
whereas the first lattice-QCD results for zero-skewness GPDs were
reported in~\cite{Chen:2019lcm,Karthik:2021sbj}. The limit of large-$N_c$ in ITDs of the pion was studied in~\cite{Karthik:2022fdb}.

The role of model studies of quasi-distributions is based not only on the fact that they are the core of lattice studies
of partonic distribution.  These quantities are interesting per se as properties of hadrons related to matrix elements of bilocal
operators. They also shed light on the rather intricate formalism of
partonic distributions, providing nontrivial examples. This is very much so in the case of the pion which arises as the would-be Goldstone boson of the spontaneously broken chiral symmetry.

In this work we study the qGPDs and pGPDs, as well as the generalized
Ioffe-time distributions of the
pion in the framework of the NJL model at the low quark model scale,
where the chiral symmetry features of QCD are properly implemented and expected to largely dominate the results. We also
consider the transversity distributions, related to the spin
distributions.  With the obtained analytic or semi-analytic
expressions we study the limit of the large momentum $P_z$ of the
pion, i.e., we investigate the approach of the quasi distributions to
the standard GPD case. The question is of practical importance for the
scheme to work, as on the lattice the upper value of $P_z$ is
naturally limited with the inverse lattice spacing, $a \sim 1/P_z$, which presently reaches (in physical units) up to about $a \sim
  0.05-0.10$~fm and, consequently, $P_z \sim 2-4 $~GeV only.  We stress
that our model approach satisfies all the general field-theoretic
requirements, such as the Lorentz covariance, the gauge invariance, or
crossing symmetry.

In this paper we restrict to the model
results at the quark model scale~\cite{Broniowski:2007si}, and do not
carry out the QCD evolution to higher scales.  We are interested in
exploring the $P_z \to \infty$ limit, which is generally not affected
by the evolution in the sense that the discrepancy between a finite
$P_z$ and the infinite limit at the quark model scale would be carried
over with the evolution to higher scales. We note that a working
scheme for the evolution of qGPD, proceeding via evolution of the
$k_T$-unintegrated
distributions~\cite{Kwiecinski:2002bx,Gawron:2002kc,Gawron:2003qg,RuizArriola:2004ui},
has been used in~\cite{Broniowski:2017gfp}. Such studies for the
present case, needed to compare the model results to the lattice data
obtained at much higher scales, are left for a future work.

We find that to obtain reasonable GPDs, in particular the
non-forward ones, requires large values of $P_z$, at least of the
order of $\sim 3$~GeV. We also explore polynomiality of the quasi
distributions and study in some detail the generalized quasi form
factors. The issue of separability of the transverse and longitudinal
dynamics in the model is studied with the help of the generalized
Ioffe-time distributions

The paper is organized as follows: In Section~\ref{sec:basics} we
review and extend the general formalism of quasi GPDs of the pion,
discussing in particular the feature of polynomiality and the
generalized quasi form factors, the generalized Ioffe-time
distributions, and the generalized pseudo-distributions, with their
relation to the $k_T$-unintegrated GPDs. Section~\ref{sec:val} is
devoted to our model results. We first briefly review the NJL model
and then present its numerous analytic or semi-analytic results for
the quasi distributions and related quantities introduced earlier. The
proximity of the results obtained at a large but finite $P_z$ to
the $P_z \to \infty$ limit is assessed. We present the generalized
Ioffe-time distributions and the generalized pseudo-distributions and
discuss the issue of the separability of the longitudinal and
transverse dynamics. We also elucidate the range in the Ioffe time needed to reliably pass to
the $x$ space. The Appendices contain some more technical but
nevertheless useful and relevant results, such as explicit
realizations of various possible kinematics for the
quasi-distributions, the derivation and explicit forms of the one-loop
expressions for the NJL model, or an explicit check of polynomiality for the
quasi distributions.

\section{Basics \label{sec:basics}}

Given the large number of variables and their Fourier conjugates
  involved in the GPDs and related objects, there is some inherent
  degree of complexity which cannot be avoided.  In this Section we
provide a concise glossary of the necessary definitions and establish the
notation of our paper. The mentioned results are general, independent
of the model used later on.

\subsection{Definitions \label{sec:def}}

The valence quark GPDs and qGPDs of the pion can be defined via the same universal formula involving bilocal quark operators~\cite{Ji:1996ek,Ji:1996nm,Radyushkin:1997ki,Radyushkin:1998es,Polyakov:1999gs} (in Appendix~\ref{app:decomp} we discuss possible more general definitions related to the Lorentz decomposition of the amplitude~\cite{Orginos:2017kos}), namely 
\begin{eqnarray}
&&    \int _{-\infty}^{\infty} \! \frac{d\lambda}{4\pi} e^{i\lambda y \, p\cdot n}\langle\pi^b(p+q)|\overline{\psi}_\alpha(-\tfrac{\lambda}{2}n)\, \slashed{n} \, \psi_\beta(\tfrac{\lambda}{2}n)|\pi^a(p)\rangle \nonumber \\ 
&&= \delta^{ab}\delta_{\alpha\beta}\mathcal{H}^{I=0}(y,\zeta,t,n^2)+i\epsilon^{abc}\tau^c_{\alpha\beta}\mathcal{H}^{I=1}(y,\zeta,t,n^2), \nonumber \\\label{eq:Hdef}
\end{eqnarray}
where $a$, $b$, and $c$ are the isospin indices, $\alpha$ and $\beta$ are the quark flavors (summation over color is implicit), and the subscripts $I=0,1$ denote the isospin. As is routinely being done in low-energy chiral quark models, the gauge link operators are omitted, as one does not consider gluons at the quark model scale. Furthermore,  the ingoing and outgoing pions are on the mass shell, $t$ is the momentum transfer, and $\zeta$ is the skewness parameter:
\begin{eqnarray}
p^2 = m_\pi^2, \;p\cdot n = 1, \; q\cdot n = -\zeta, \; q^2 = -2p\cdot q = t.  \label{eq:momcond}
\end{eqnarray}
The momentum fraction carried by the quark is denoted as
\begin{eqnarray}
y = \frac{k\cdot n}{p\cdot n}.
\end{eqnarray}
For the GPDs, the quarks are separated by a light-like (null) vector
$\lambda n$, i.e., $n^2=0$ (then $y$ is traditionally written as the Bjorken $x$),
whereas for qGPDs we have a space-like separation, with $n^2 < 0$, in certain harmony with the Euclidean nature of these objects on the lattice.

For the valence transversity case, tGPD and qtGPD are defined via an analogous formula to Eq.~(\ref{eq:Hdef}), with an additional tensorial structure pulled out:
\begin{eqnarray}
&&     \int  \! \frac{d\lambda \,e^{i\lambda y \, p\cdot n}}{4\pi}\langle\pi^b(p\!+\!q)|\overline{\psi}_\alpha(-\tfrac{\lambda}{2}n)  n_\mu \sigma^{\mu\nu}\gamma_5 \psi_\beta(\tfrac{\lambda}{2}n)|\pi^a(p)\rangle \nonumber \\ 
&& \;\; = \epsilon^{n p q \nu}  \left[ \delta^{ab}\delta_{\alpha\beta}\mathcal{E}^{I=0}(y,\zeta,t,n^2) \right .\nonumber \\
&& \hspace{1cm}+ \left . i\epsilon^{abc}\tau^c_{\alpha\beta}\mathcal{E}^{I=1}(y,\zeta,t,n^2) \right], \nonumber \\ \label{eq:Edef}
\end{eqnarray}
where $\epsilon^{n p q \nu}=\epsilon^{\rho\sigma\lambda\nu}n_\rho p_\sigma q_\lambda$ involves the Levi-Civita tensor with the convention $\epsilon^{0123}=-1$. Note that this definition leads to $\mathcal{E}^{I=0,1}$ of dimension of inverse mass\footnote{This can be compensated by multiplying with the mass of the hadron, but we chose not to do it, as it is not usable for the pion in the chiral limit. Also, physical observables involve the whole matrix element, where the issue does not arise.}, while $\mathcal{H}^{I=0,1}$ are dimensionless. For the GPD or tGPD case, Eqs.~(\ref{eq:Hdef},\ref{eq:Edef}) define the leading twist-2 distributions in the pion.

The definitions (\ref{eq:Hdef}) and (\ref{eq:Edef}) are fully relativistically 
covariant, and so is our evaluation. However, in
Appendix~\ref{app:kin} we provide some possible explicit kinematic
assignments to the vectors $p$, $q$, and $n$, defining various
reference frames. In particular, for the kinematics used by
Ji~\cite{Ji:2013dva}, where the pion moves with momentum $P_z$ and
$n=(0,0,0,-1/P_z)$, one has
\begin{eqnarray}
n^2=-\frac{1}{P_z^2}. \label{eq:nPz}
\end{eqnarray}

With the skewness $1\ge\zeta\ge 0$, and when $n^2=0$ (the GPD and tGPD cases), the momentum fraction is bounded $x\in[-1+\zeta,1]$ and has three distinct sub-domains: $[-1+\zeta,0]$, $[0,\zeta]$, and $[\zeta,1]$, corresponding to three distinct virtual processes: excitement and de-excitement of the antiquark, excitement of a quark-antiquark pair, and excitement and de-excitement of a quark, respectively. On the contrary, for $n^2<0$ (the qGPD and qtGPD cases), 
the momentum fraction $y$ is unbounded, $y\in (-\infty,\infty)$, and the above-mentioned kinematic regions are not sharply separated. 

\subsection{Asymmetric and symmetric conventions}

Two conventions are used in the literature for the momentum fractions ($y$, $Y$) and the skewness ($\zeta$, $\xi$), which can be determined relative to the incident pion momentum, or to the average momentum of the incident and outgoing pions. These two sets are related by
\begin{align}
    Y = \frac{2y-\zeta}{2-\zeta}, \;\;\;  \xi = \frac{\zeta}{2-\zeta}. \label{eq:Yy}
\end{align}
The variables $Y$ and $\xi$ correspond the so-called symmetric convention, because of the symmetry or antisymmetry of the qGPDs about $Y=0$, 
whereas $y$ and $\zeta$ relate to the asymmetric convention. In this paper we use the symbols $H$ and $E$ to represent qGPDs and tqGPDs in the symmetric convention, while $\mathcal{H}$ and $\mathcal{E}$ are used for the asymmetric convention. Switching between the two conventions amounts to the replacement 
\begin{eqnarray}
H^{I=0,1}(Y,\xi,t,n^2)&=& \mathcal{H}^{I=0,1}(y,\zeta,t,n^2), \label{eq:passH} \\ 
E^{I=0,1}(Y,\xi,t,n^2)&=&\frac{1}{1+\xi}\mathcal{E}^{I=0,1}(y,\zeta,t,n^2). \label{eq:pass}
\end{eqnarray}
We note that according to definition~(\ref{eq:Hdef}) the normalization is $\int dy\, \mathcal{H}^{I=1}(y,\zeta,t=0,n^2)=1+\zeta/2$, hence $\mathcal{H}^{I=1}$ is not normalized to 1 (the conserved charge of the pion). That feature could be enforced by altering Eq.~(\ref{eq:Hdef}), putting instead of $1/(4\pi)$ the factor $1/[2\pi (2-\zeta)]$. Then that factor would be canceled in Eq.~(\ref{eq:passH}) by the Jacobian of the transformation from $y$ to $Y$, which is \mbox{$dy/dY= 2-\zeta=1/(1+\xi)$}. The traditional convention, however, uses Eqs.~(\ref{eq:Hdef},\ref{eq:passH}) as written, with $\int dY\, H^{I=0,1}(Y,\xi,t=0,n^2)=1$. For $E^{I=0,1}$ there is no conservation law enforcing normalization. Nevertheless, with definition (\ref{eq:Edef}) and the Jacobian kept in Eq.~(\ref{eq:pass}) the normalization of $E^{I=0,1}(Y,\xi,t=0,n^2)$ is independent of $\xi$. Further discussion of related issues is given in subsection~\ref{sec:gff}.

We will occasionally use a short-hand notation, where 
$\mathcal{F}$ stands for  $\mathcal{H}^{I=0,1}$ or $\mathcal{E}^{I=0,1}$, and similarly $F$ stands for $H^{I=0,1}$ or $E^{I=0,1}$.
The basic symmetry features of qGPDs in the symmetric notation are
\begin{eqnarray}
F^{I=0}(Y,\xi,t,n^2)&=&-F^{I=0}(-Y,\xi,t,n^2), \nonumber \\
F^{I=1}(Y,\xi,t,n^2)&=&F^{I=1}(-Y,\xi,t,n^2).
\end{eqnarray}
The corresponding quark and antiquark qGPDs and \mbox{qtGPDs} are constructed with the relations
\begin{align}
    \mathcal{F}_q(y,\zeta,t,n^2)&=\frac{1}{2}\left(\mathcal{F}_{I=0}(y,\zeta,t,n^2)+\mathcal{F}_{I=1}(y,\zeta,t,n^2) \right), \nonumber\\
    \mathcal{F}_{\bar{q}}(y,\zeta,t,n^2)&=\frac{1}{2}\left(\mathcal{F}_{I=0}(y,\zeta,t,n^2)-\mathcal{F}_{I=1}(y,\zeta,t,n^2) \right), \label{eqqbgpd}
\end{align}
and analogously for the symmetric notation. 

For GPDs and tGPDs, $x\in [-1+\zeta,0] \cup [\zeta,1]$, or $X \in
[-1,-\xi] \cup [\xi,1]$, is referred to as the DGLAP region, whereas
$x\in [0,\zeta]$, or $X \in [-\xi,\xi]$, as the ERBL region, with the
names borrowed from the corresponding
Dokshitzer-Gribov-Lipatov-Altarelli-Parisi QCD evolution equations for
the PDFs and the Efremov-Radyshkin-Brodsky-Lepage evolution for the
DAs~\cite{Efremov:1979qk,Lepage:1980fj}.

\subsection{Polynomiality \label{sec:polyno}}

An important formal property of qGPDs is the polynomiality feature, whereby the $Y$-moments of $F$ are even polynomials in the $\xi$ variable, with the coefficients $A^{(k)}_m$ and $B^{(k)}_m$, depending on $t$ and $n^2$, interpreted as the generalized quasi form factors:
\begin{eqnarray}
\int_{-\infty}^\infty dY \, Y^{2k} H^{I=1}(Y,\xi,t,n^2) &=& \sum_{m=0}^k A^{(k)}_m(t,n^2) \xi^{2m},\nonumber \\
\int_{-\infty}^\infty dY\,  Y^{2k+1} H^{I=0}(Y,\xi,t,n^2) &=& \sum_{m=0}^{k+1} B^{(k)}_m(t,n^2) \xi^{2m}. \nonumber \\  
\int_{-\infty}^\infty dY \, Y^{2k} E^{I=1}(Y,\xi,t,n^2) &=& \sum_{m=0}^k A^{(k)}_{Tm}(t,n^2) \xi^{2m},\nonumber \\
\int_{-\infty}^\infty dY\,  Y^{2k+1} E^{I=0}(Y,\xi,t,n^2) &=& \sum_{m=0}^{k+1} B^{(k)}_{Tm}(t,n^2) \xi^{2m}.  \nonumber \\
\label{eq:pol}
\end{eqnarray}
Polynomiality follows from the basic field theoretic features of the theory, such as the Lorentz covariance, time reversal, and hermiticity, in a full analogy to the GPD case ($n^2=0$). The simple derivation proceeds via the double distributions~\cite{Radyushkin:1996nd}, which now are additionally functions of $n^2$.
This $n^2$ dependence is carried over to the form factors.

In Appendix~\ref{app:poly} we explicitly check polynomiality for the basic one-loop functions entering the NJL evaluation. Clearly, in the limit of $n^2 \to 0$ the generalized quasi form factors $A^{(k)}_m(t,n^2)$ and $B^{(k)}_m(t,n^2)$ tend to the generalized form factors related to the GPDs (for a detailed discussion of these quantities and their QCD evolution see~\cite{Broniowski:2008hx}). 

The zeroth moment of $H^{I=1}$ is independent of $n^2$,
\begin{eqnarray}
\int_{-\infty}^\infty \!\!\! dY H^{I=1}(Y,\xi,t,n^2)= A^{(0)}_0(t)=2 F_V(t), \label{eq:FV}
\end{eqnarray}
where $F_V(t)$ is the pion's charge form factor. 
Similarly,
\begin{eqnarray}
\int_{-\infty}^\infty \!\!\! dY E^{I=1}(Y,\xi,t,n^2)= A^{(0)}_{T0}(t), \label{eq:E1} \\
\int_{-\infty}^\infty \!\!\! dY E^{I=0}(Y,\xi,t,n^2)= B^{(0)}_{T0}(t),
\end{eqnarray}
are independent of $n^2$.
The higher moments do depend explicitly on $n^2$.\footnote{\label{foo:mom} We note, however, that the higher moments in $Y$ may not exist due to the long-range tails in $Y$ of the $F(Y,\xi,t,n^2)$ distributions. As a mater of fact, in the employed NJL model with two Pauli-Villars subtractions we may compute, depending on the type of distribution, the first few $Y$-moments according to Eq.~(\ref{eq:asympt}).}

The first moment of $H^{I=0}$ involves the gravitational form factors $\theta_{1,2}$~\cite{Broniowski:2017gfp}:
\begin{eqnarray}
\!\!\!\!\! \int_{-\infty}^\infty \!\!\!\!\!\! dY Y H^{I=0}(Y,\xi,t,n^2)= \theta_2(t,n^2)-\xi^2 \theta_1(t,n^2). \label{eq:gr}
\end{eqnarray}
At the quark model scale, where the quarks are the only degrees of freedom, one has
$\theta_2(t=0,n^2=0)=1$, which reflects the fact that the trace of the energy-momentum tensor in the pion state is $m_\pi^2$~\cite{Broniowski:2017gfp} (the energy-momentum sum rule).
In the chiral limit, a low-energy theorem~\cite{Donoghue:1991qv} yields a general relation $\theta_2(t,n^2=0)- \theta_1(t,n^2=0)={\cal O}(m_\pi^2)$.
We note that in the mechanistic interpretation $\theta_1(t)$ is the $D$-term~\cite{Polyakov:2018zvc}.

\subsection{Generalized Ioffe-time distributions \label{sec:iof}} 

The generalized ITDs are obtained from qGPDs or \mbox{qtGPDs} via the Fourier transform from the momentum-fraction variables  to the Ioffe time $\nu$:
\begin{eqnarray}
{\cal F}_I(-\nu,\zeta,t,-z^2)&=&\int dy\, e^{i \nu y }{\cal F}(y,\zeta,t,z^2/\nu^2), \nonumber\\
{F}_I(-\nu,\xi,t,-z^2)&=&\int dY\, e^{i \nu Y }{F}(Y,\xi,t,z^2/\nu^2), \label{eq:defiof}
\end{eqnarray}
where the subscript $I$ stands for ``Ioffe" and, as already
  mentioned, $\mathcal{F}$ stands either for $\mathcal{H}^{I=0,1}$ or
  $\mathcal{E}^{I=0,1}$, while $F$ either for $H^{I=0,1}$ or
  $E^{I=0,1}$. The Ioffe time is defined as
\begin{eqnarray}
\nu=p\cdot z, \label{eq:itime}
\end{eqnarray}
which in the employed kinematics is $\nu=-P_z z_3$, hence $\nu^2=P_z^2 z_3^2=z^2/n^2$.
The variables $\nu$ and $z^2\le 0$ are the arguments of ITDs. An obvious relation between between the asymmetric and symmetric notation follows from Eq.~(\ref{eq:Yy}), namely
\begin{eqnarray}
{F}_I(-\nu,\xi,t,-z^2)=e^{-i \nu \xi} {\cal F}_I \left[ -(1+\xi)\nu,\tfrac{2\xi}{1+\xi},t,-z^2 \right ],\nonumber \\  \label{eq:mix}
\end{eqnarray}
hence for non-zero $\xi$ the real and imaginary parts of the two conventions mix, while $\nu$ gets rescaled.

Another simple fact is that the subsequent derivatives of ${F}_I(-\nu,\xi,t,-z^2)$ with respect to $\nu$ at $\nu \to 0$ (with $z^2=\nu^2 n^2$) provide the moments of Eqs.~(\ref{eq:pol}). We wish to underline here that although higher moments of \mbox{qGPDs} and \mbox{tqGPD} may not and, in general, do not exist, as remarked in footnote${}^{\ref{foo:mom}}$ and elaborated in Appendix~\ref{app:poly}, the corresponding ITDs are nevertheless well defined, as a Fourier transform of a regular function that asymptotically goes to zero always exists. The point is, however, that one cannot use it to obtain the higher (non-existing) moments, as one cannot interchange the limit of taking the derivatives and the improper integration. The problem disappears for the GPD limit, as the finite support in $x$ make all moments exist.

With the notation (\ref{eq:nPz}) we have (interpreting $P_z$ as fixed)
\begin{eqnarray}
\hspace{-5mm} {F}_I(-\nu,\xi,t,\nu^2/P_z^2)=\int dY\, e^{i \nu Y }{F}(Y,\xi,t,-1/P_z^2). \label{eq:defiofPz}
\end{eqnarray}
Taking $z=(0,0,0,z_3)$, whereby $\nu=- z_3 P_z$, one can also write
\begin{eqnarray}
\hspace{-5mm} {F}_I(z_3 P_z,\xi,t,z_3^2)=\int dY\, e^{-i z_3 P_z Y }{F}(Y,\xi,t,-1/P_z^2) \nonumber \\ \label{eq:defiofz3}
\end{eqnarray}
and treat the ITDs as functions of $z_3$ at fixed $P_z$.
Due to the rotational invariance, $z_3^2$ should be viewed as the modulus squared of the space part of $z$, namely $|{\bm z}|^2$. In particular, one can choose ${\bm z}$ in the transverse direction, where it is related to the $k_T$-unintegrated distributions via Fourier transform.\par

The generalized {\em reduced} Ioffe-time distributions~\cite{Orginos:2017kos} are defined as 
\begin{eqnarray}
\mathfrak{F}(-\nu,\xi,t,-z^2)&=&\frac{{F}_I(-\nu,\xi,t,-z^2)}{{F}_I(0,\xi,t,-z^2)} \nonumber \\
&=&\frac{\int dY\, e^{i \nu Y }{F}(Y,\xi,t,z^2/\nu^2)}{\int dY\,{F}(Y,\xi,t,z^2/\nu^2)}, \label{eq:redI}
\end{eqnarray}
or
\begin{eqnarray}
\hspace{-5mm} \mathfrak{F}(z_3 P_z,\xi,t,z_3^2)=\frac{\int dY\, e^{-i z_3 P_z Y }{F}(Y,\xi,t,-1/P_z^2)}{\int dY\,{F}(Y,\xi,t,-1/P_z^2)}. \label{eq:redI2}
\end{eqnarray}
These quantities can be efficiently used to probe the transverse-longitudinal factorization~\cite{Orginos:2017kos,Broniowski:2017gfp}. Their $Y$-moments are independent of $P_z$ up to the rank 1 (see the discussion in subsection~\ref{sec:gff}). Moreover, the reduced distributions are advantageous for the lattice QCD simulations~\cite{Orginos:2017kos}.

\subsection{Generalized pseudo-distributions \label{sec:pse}} 

Radyushkin's pseudo-distributions~\cite{Radyushkin:2016hsy,Radyushkin:2017cyf,Radyushkin:2017lvu,Radyushkin:2017gjd} are in turn defined as Fourier transforms of the ITDs from $\nu$ to the momentum fraction $x$, namely
\begin{eqnarray}
{F}_P(x,\xi,t,-z^2)=\int \frac{d\nu}{2\pi} \, e^{-i \nu x }{F}_I(-\nu,\xi,t,-z^2) \label{eq:psdef}
\end{eqnarray}
(and similarly for the asymmetric convention),
where the superscript $P$ stand for ``pseudo". An advantage of these distributions is that the momentum fraction has the support $x \in [-1,1]$ (cf. Appendix~\ref{app:ioffe}). Moreover, the $x \in [0,1]$ range corresponding to the quarks is strictly separated from the  $x \in [-1,0]$  region corresponding to the antiquarks. The functional dependence on $x$ and $z^2$ in the generalized pseudo-distributions can be used to probe the correlation between the longitudinal and transverse dynamics, since in the case of no correlations the $x$ and $z^2$ dependence factorizes. 
Clearly, from the definitions it follows that ${F}_P(x,\xi,t,-z^2=0)={F}(x,\xi,t,n^2=0)$.

\begin{figure}[tb]
\centering
\includegraphics[scale=0.42]{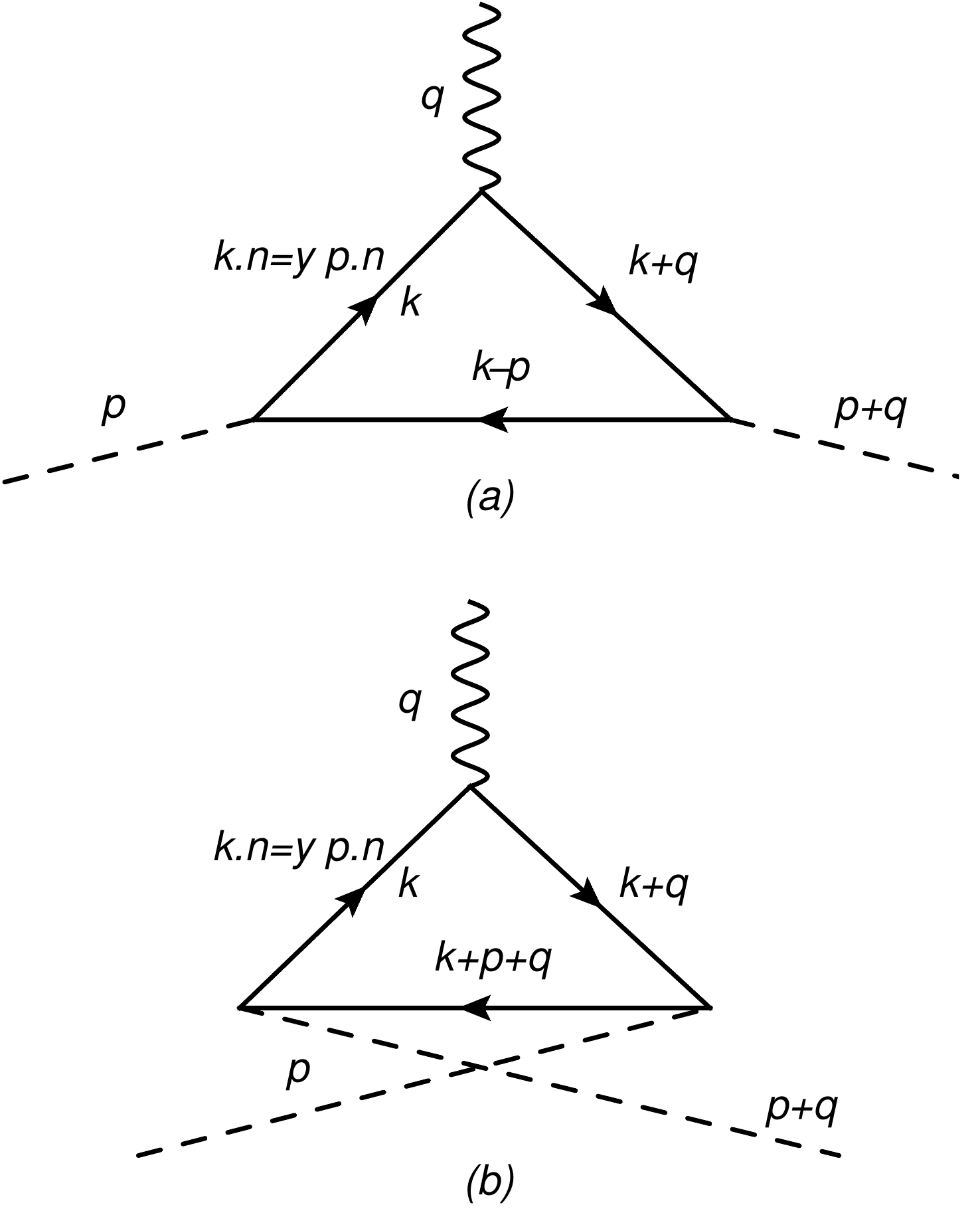} \\
\vspace{3mm}
\includegraphics[scale=0.42]{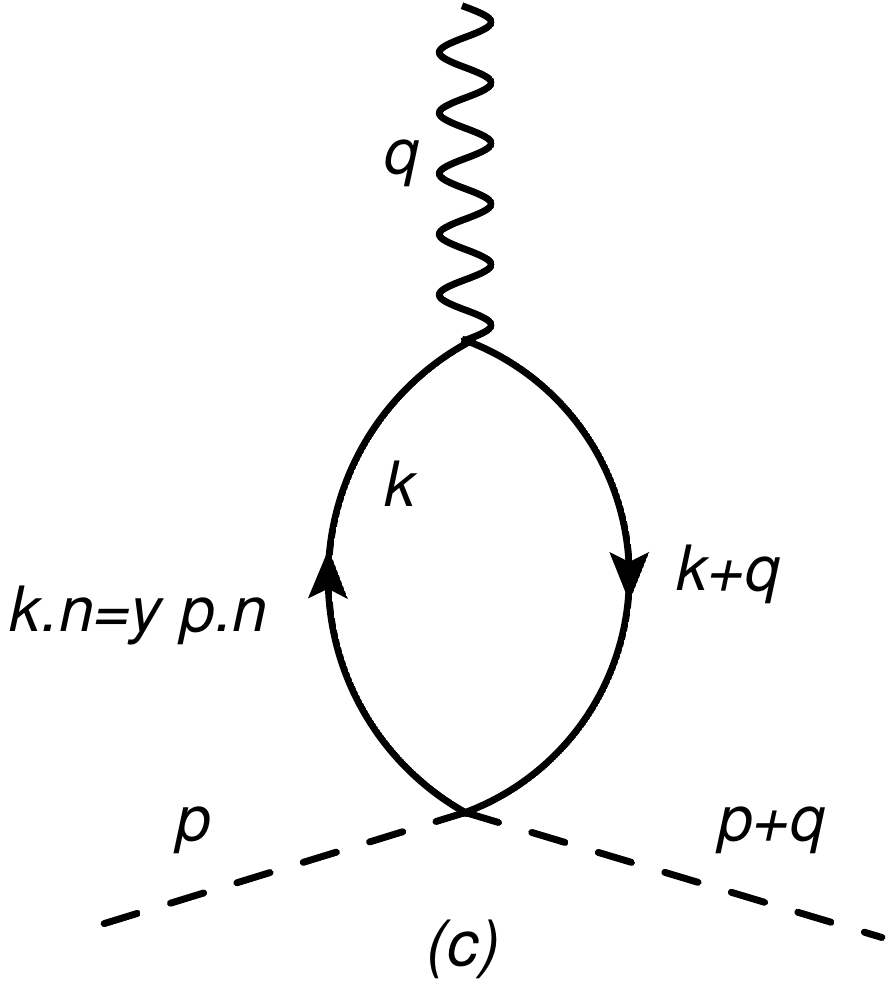}
\caption{One-loop diagrams for the evaluation of qGPDs and qtGPDs in the NJL model. Solid lines indicate the quark propagator and dashed lines the external on-shell pions. The wavy line denotes  the probing operator $\slashed{n}$ or $n_\mu \sigma^{\mu \nu} \gamma_5$ for \mbox{qGPDs} or qtGPDs, respectively. The loop momentum integration is over three dimensions only, with the constraint $k\cdot n=y \, p \cdot n$ eliminating the fourth. \label{fig:feyndia}}
\end{figure}

\begin{figure*}[tb]
\includegraphics[width=0.43\textwidth]{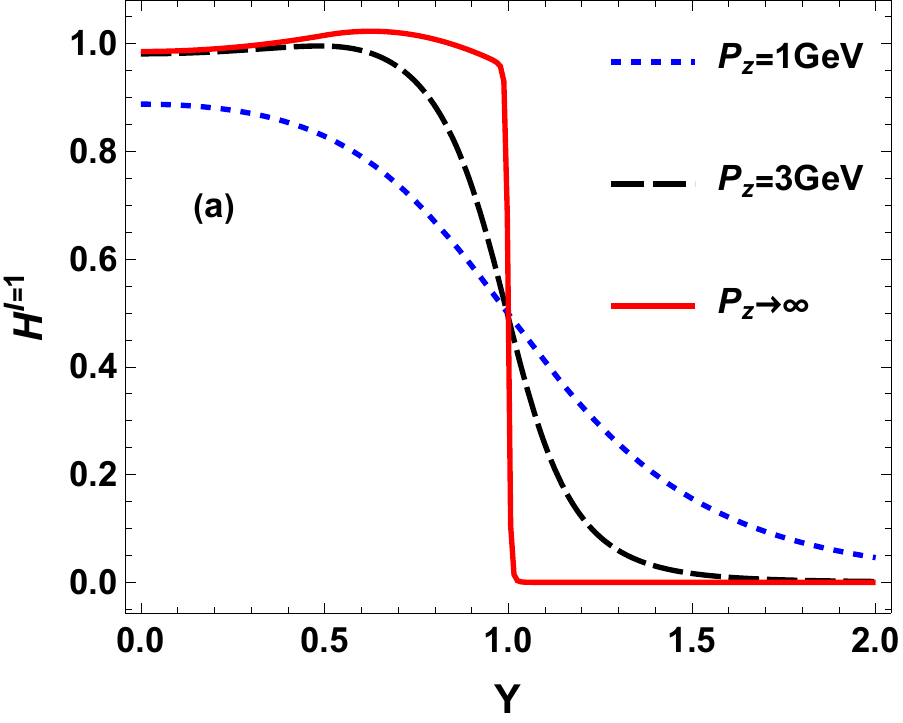} \hfill \includegraphics[width=0.43\textwidth]{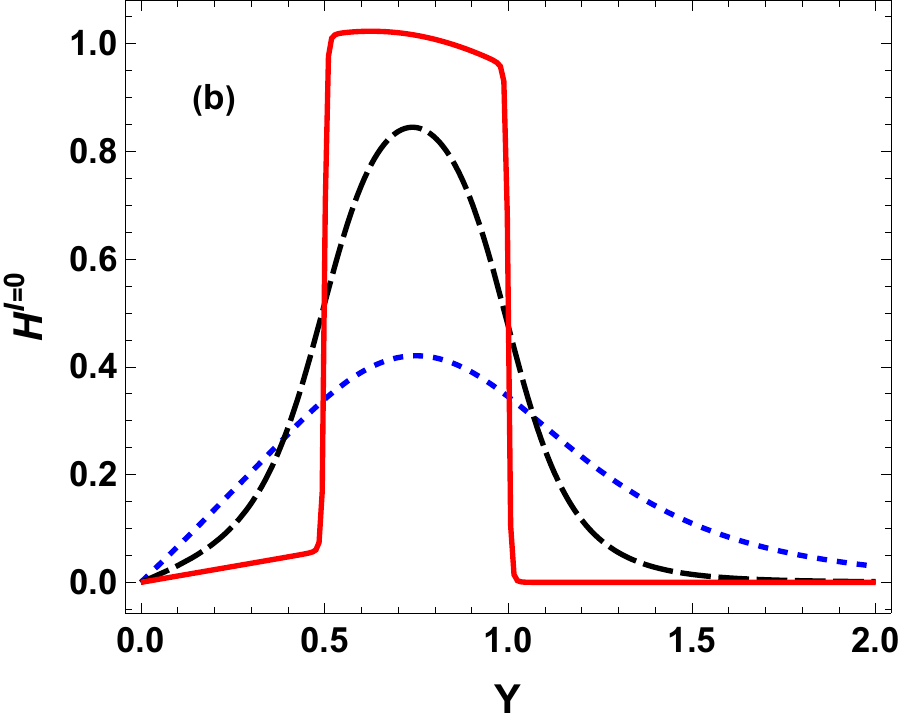} \\ ~ \\
\includegraphics[width=0.43\textwidth]{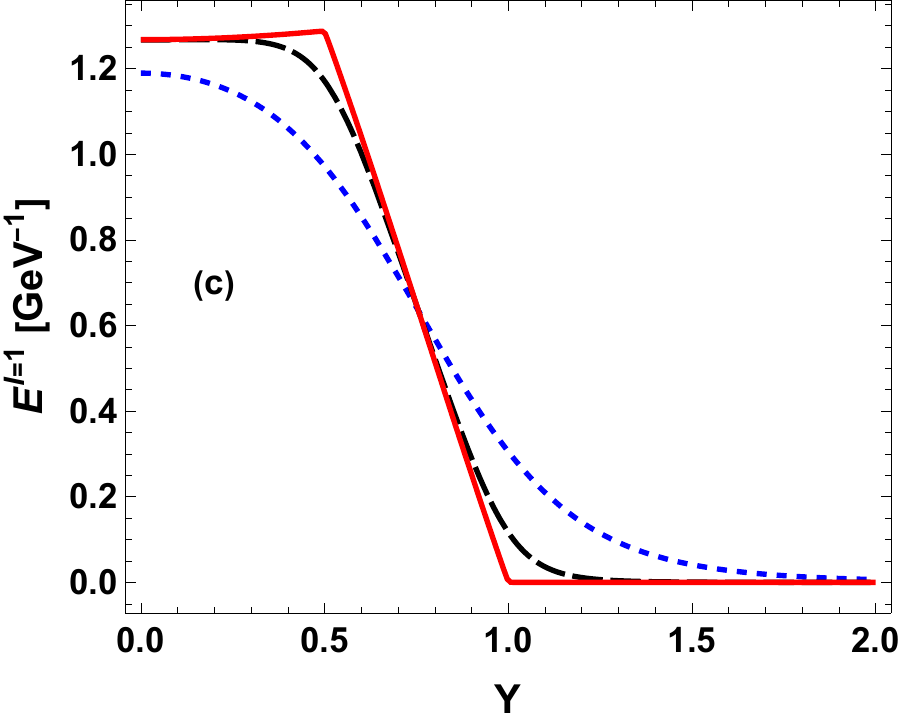} \hfill \includegraphics[width=0.43\textwidth]{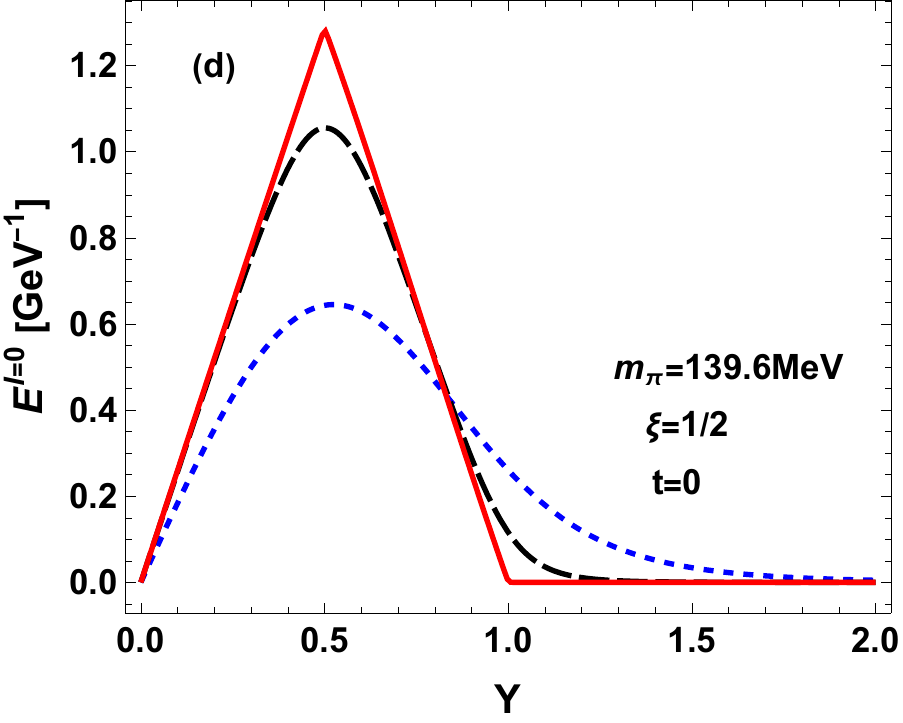} 
\caption{qGPDs $H^{I=0,1}$ and qtGPDs $E^{I=0,1}$, plotted as functions of $Y$ for several values of $P_z$ (physical pion mass, $\xi=1/2$, and  $t=0$).\label{fig:allPz}}
\end{figure*}

\subsection{Relation to $k_T$-unintegrated GPDs}

Another property of pGPDs is their simple relation to the $k_T$-unintegrated GPDs (or the transverse-momentum distributions (TMDs), up to the intricacies of the Wilson link operators~\cite{Constantinou:2017sej}). This feature, derived by Radyushkin~\cite{Radyushkin:2016hsy,Radyushkin:2017cyf,Radyushkin:2017lvu,Radyushkin:2017gjd} and following solely from the Lorentz covariance, naturally generalizes to the pGPD case. The $k_T$-unintegrated GPDs can be defined as a Fourier transform of the generalized pseudo-distributions as follows:
\begin{eqnarray}
&& F_T(x,\xi,t,k_1^2+k_2^2)= \label{eq:kT}\\
&&\hspace{1cm} \int \frac{dz_1 dz_2}{(2\pi)^2} e^{ik_1 z_1+ik_2 z_2} F_P(x,\xi,t,z_1^2+z_2^2),\nonumber
\end{eqnarray}
where indices $1,2$ relate to the transverse space. The generalization of the Radyushkin relation, written invariantly, reads
\begin{eqnarray}
&& F(Y,\xi,t,n^2)= \label{eq:radrel} \\
&&~~~\frac{1}{\sqrt{-n^2}} \int dk_1 \int dx\, F_T\left[x,\xi,t,k_1^2-\frac{(x-Y)^2}{n^2} \right]. \nonumber 
\end{eqnarray}
An explicit check for the one-loop calculation is provided in Appendix~\ref{app:rad}.

The relation of the generalized pseudo or quasi distributions to the $k_T$-unintegrated generalized distributions opens the possibility of carrying out the QCD evolution with the approach suggested by Kwieci\'nski~\cite{Kwiecinski:2002bx,Gawron:2002kc,Gawron:2003qg,RuizArriola:2004ui}, as already done for the qGPDs of both the pion and the nucleon in~\cite{Broniowski:2017gfp}. This interesting problem is left for a future study.

\begin{figure*}[tb]
\includegraphics[width=0.43\textwidth]{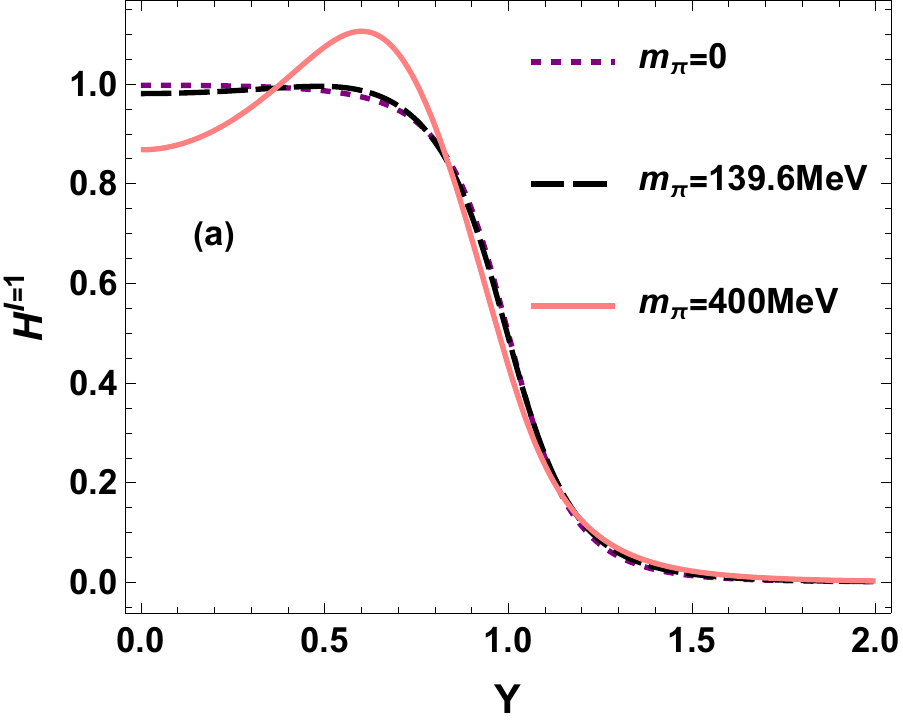} \hfill \includegraphics[width=0.43\textwidth]{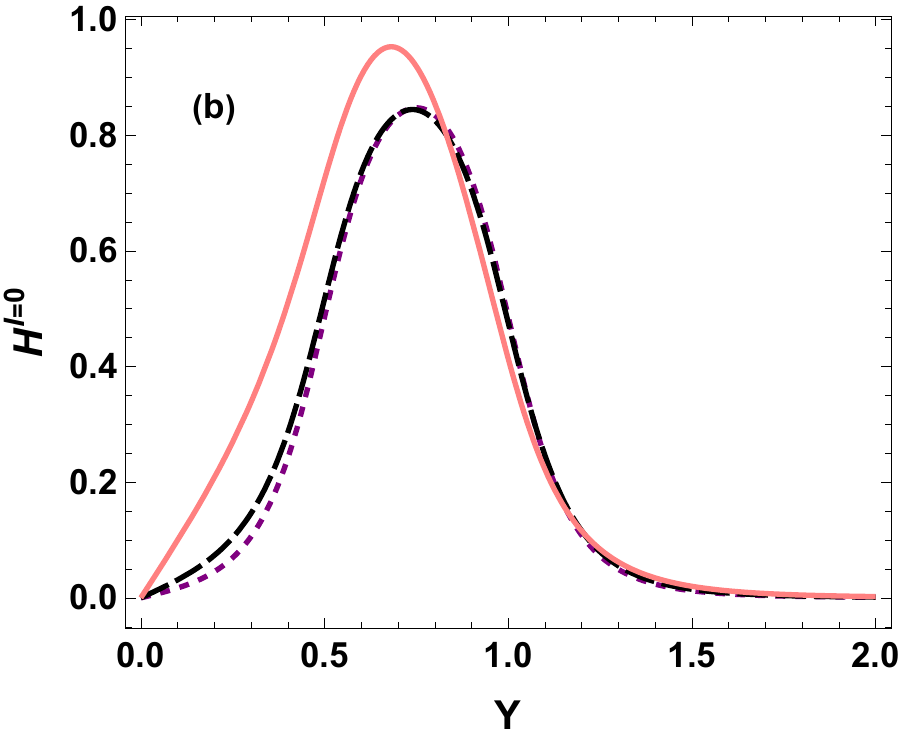} \\ ~ \\
\includegraphics[width=0.43\textwidth]{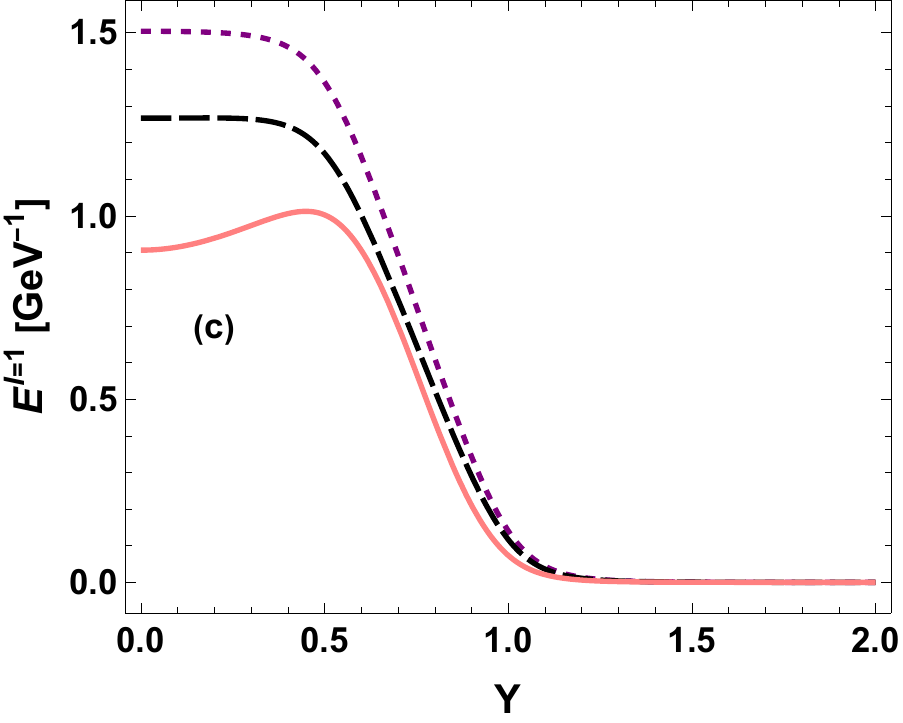} \hfill \includegraphics[width=0.43\textwidth]{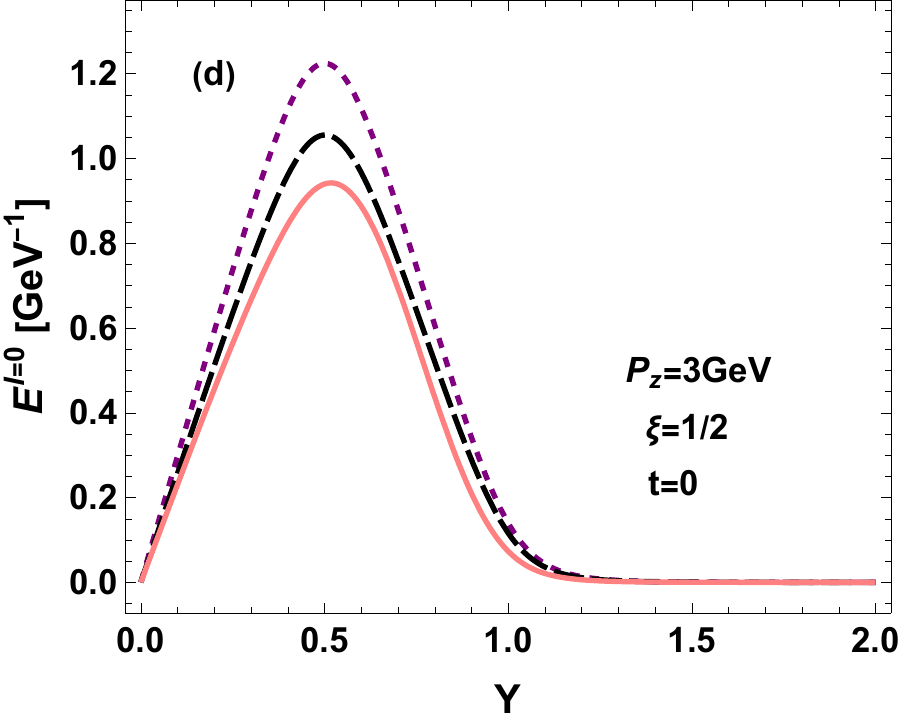} 
\caption{Same as in Fig.~\ref{fig:allPz}, but for a varying pion mass and fixed $P_z = 3$~GeV, $\xi=1/2$, and $t=0$. \label{fig:allmpi}}
\end{figure*}

\section{Valence GPDs and tGPDs of the pion in the NJL model \label{sec:val}}

This Section presents novel analytic and semi-analytic results, specific to the NJL model in the large-$N_c$ limit, i.e., obtained at the one-quark-loop level. All the discussion concerns the valence distributions at the quark model scale. 

\subsection{Lagrangian}

In this subsection we describe the calculation and present the results for the valence qGPDs and qtGPDs of the pion in the NJL model. We apply the non-linear 
version of the model with the Lagrangian of the form 
\begin{eqnarray}
{\cal L}=\overline{\psi} \left ( i\slashed{\partial } - M U^5 - m \right )  \psi, \label{eq:lag}
\end{eqnarray}
with
\begin{eqnarray}
U^5=\exp \left( i \gamma_5 {\bm \tau}\cdot{\bm \phi}/f \right) , \label{eq:U5}
\end{eqnarray}
where $M$ is the constituent quarks mass from the spontaneous chiral symmetry breaking, $m$ is the current quark mass explicitly breaking the chiral symmetry, $f$ is the pion weak decay constant, ${\bm \tau}$ are the Pauli matrices, and ${\rm \phi}$ is the pion field. When departing from the strict chiral limit, the canonical pion field becomes ${\rm \pi}=Z {\rm \phi}$, such that the canonical pion-quark coupling constant obtained from the residue 
at the pole of the pion propagator is $g_{\pi q}=ZM/f$ (see~\cite{RuizArriola:2002wr} for details).

\subsection{Methodology}

Our methodology follows the numerous earlier works on the properties of the pion: 
PDF~\cite{Davidson:1994uv,Davidson:2001cc,Weigel:1999pc}, the distribution
amplitude~\cite{RuizArriola:2002bp}, the generalized distribution
functions~\cite{Broniowski:2007si}, the generalized form factors~\cite{Broniowski:2008hx}, the quasi distribution
amplitude~\cite{Broniowski:2017wbr}, the quasi or
pseudo PDFs~\cite{Broniowski:2017gfp}, as well as the double
distribution functions~\cite{Broniowski:2019rmu}.  
The calculation at the large-$N_c$ level amounts to the evaluation of one-loop diagrams displayed in the Fig.~\ref{fig:feyndia}. We note that diagram (c), appearing in the nonlinear model~(\ref{eq:lag}) for nonzero $q$, makes the result covariant and contributes to the $D$-term~\cite{Polyakov:1999gs}.\footnote{In the linear $\sigma$-like model following from the bosonization of the NJL model, the diagram equivalent to Fig.~\ref{fig:feyndia}(c) involves a $\sigma$-meson propagator in the $t$-channel. The results at non-zero $t$ are somewhat more complicated from the present ones, but the essentials are the same.}   

The results are interpreted as pertaining to the {\em quark-model
  scale}, where the quarks are the only dynamical degrees of
freedom. This scale is very low, $\sim 320$~MeV, as can be assessed
from the momentum fraction carried by the valence
quarks~\cite{Broniowski:2007si}, as well as independently from the
value of the quark condensate~\cite{Broniowski:2021awb}. There is a
demand for the QCD evolution whenever one wishes to compare the model
results to the experimental or lattice data at much higher scales. The
results at the quark model scale, as those presented in this work, are
treated as initial conditions for the QCD evolution, with only valence
quarks present. One carries out the evolution to a higher scale, say
2~GeV, whereby sea quarks and gluons are radiatively generated. The
program has been carried out for the
PDFs~\cite{Davidson:1994uv,Davidson:2001cc},
GPDs~\cite{Broniowski:2007si}, tGPDs~\cite{Dorokhov:2011ew}, and
qPDFs~\cite{Broniowski:2017gfp}. For the present case of qGPDs and
tqGPDs, it is left for a future study.

The NJL model is a non-renormalizable effective field-theoretical Lagrangian, hence a high-energy cut-off must be
introduced. A way to do it consistently, with the Lorentz and gauge
symmetries preserved, is to use a twice-subtracted Pauli-Villars (PV)
regularization \cite{RuizArriola:2002wr}. In this set up, the basic
scalar loop integrals $L(M^2)$ are regularized in the following way:
\begin{align}
L(M^2)_{\rm reg} &= L(M^2)-L(M^2+\Lambda^2) + \Lambda^2\frac{dL(M^2+\Lambda^2)}{d\Lambda^2}. \label{eq:pvr}
\end{align}

In deriving the one-loop NJL expressions for the \mbox{qGPDs} and tqGPDs we
follow exactly the same steps as given in
Ref. \cite{Broniowski:2007si}. Some details are provided in
Appendix~\ref{app:loops}. The amplitudes obtained from the subsequent diagrams of
Fig. \ref{fig:feyndia} are
\begin{eqnarray}
&& \mathcal{H}^a(y,\zeta,t,n^2) = \frac{i N_c g_{\pi q}^2}{4\pi^4}\int d^4k \frac{\delta(k\cdot n-y)}{D_k D_{k+q}D_{k-p}} \times \nonumber \\
&& \; \left[ (k^2-M^2)(\zeta-y-1) + k\cdot p(2y-\zeta) -k\cdot q -\tfrac{1}{2}yt \right], \nonumber \\
&&    \mathcal{H}^b(y,\zeta,t,n^2) = \frac{i N_c g_{\pi q}^2}{4\pi^4}\int d^4k \frac{\delta(k\cdot n-y)}{D_k D_{k+q}D_{k+p+q}} \times \nonumber \\
&& \; \left[ (k^2-M^2)(1-y) - k\cdot p(2y-\zeta) + k\cdot q (1-2y -\tfrac{1}{2}yt \right], \nonumber \\
&&    \mathcal{H}^c(y,\zeta,t,n^2) = \frac{i N_c g_{\pi q}^2}{4\pi^4}\int d^4k \frac{\delta(k\cdot n-y)}{D_k D_{k+q}}\left( 2y-\zeta \right), \label{eq:1lgen}
\end{eqnarray}
where, $D_\ell = \ell^2-M^2 +i0$, and the superscripts correspond to the labels in Fig. \ref{fig:feyndia}. For the proper isospin combinations one has
\begin{align}
    \mathcal{H}^{I=0} &=\mathcal{H}_a+\mathcal{H}_b+\mathcal{H}_c, \nonumber \\
    \mathcal{H}^{I=1} &= \mathcal{H}_a-\mathcal{H}_b, \label{eq:hh}
\end{align}
and explicitly
\begin{eqnarray}
&&    \mathcal{H}^{I=0,1}(y,\zeta,t,n^2) = \frac{-i N_c g_{\pi q}^2}{8\pi^4} \int d^4k \delta(k\cdot n-y) \label{Heq}\\
&&    \left( \frac{1}{D_k D_{k-p}} + \frac{1-\zeta}{D_{k+q} D_{k-p}} \mp \frac{1}{D_{k+q}D_{k+q+p}}\mp \frac{1-\zeta}{D_k D_{k+p+q}} \right.\nonumber\\
&&    \left. +\frac{(\zeta-2y)m_\pi^2 +t(y-1)}{D_k D_{k+q}D_{k-p}}\mp \frac{(\zeta-2y)m_\pi^2+t(y-\zeta+1)}{D_k D_{k+q}D_{k+p+q}} \right).\nonumber
\end{eqnarray}
Analogously, the amplitudes corresponding to the tGPDs are
\begin{eqnarray}
&&\mathcal{E}^a(y,\zeta,t,n^2) = \frac{-i N_c M g_{\pi q}^2}{4\pi^4}\int d^4k \frac{\delta(k\cdot n-y)}{D_k D_{k+q}D_{k-p}}, \nonumber \\
&&\mathcal{E}^b(y,\zeta,t,n^2) = \frac{i N_c M g_{\pi q}^2}{4\pi^4}\int d^4k \frac{\delta(k\cdot n-y)}{D_k D_{k+q}D_{k+p+q}}, \nonumber \\
&&\mathcal{E}^c(y,\zeta,t,n^2) = 0,
\end{eqnarray}
Thus, tGPDs get contributions only from the triangle diagrams of Fig.~\ref{fig:feyndia}. For the isospin combinations one has, in analogy to Eq.~(\ref{eq:hh}),
\begin{align}
   & \mathcal{E}^{I=0,1}(y,\zeta,t,n^2) = \frac{-i N_c M g_{\pi q}^2}{4\pi^4} \int d^4k \delta(k\cdot n-y)\nonumber\\
    &\left(\frac{1}{D_k D_{k+q}D_{k-p}}\mp \frac{1}{D_k D_{k+q}D_{k+p+q}} \right).\label{ETeq}
\end{align}

From the form of Eqs.~(\ref{Heq}) and (\ref{ETeq}) it is clear that we need to evaluate two types of integrals: the scalar two-point function (bubble) $I$ and the three-point function (triangle) $J$ (see the Appendix~\ref{app:loops} for the definitions and evaluation).
Substituting these in Eqs.~(\ref{Heq}) and (\ref{ETeq}) yields our final expressions used for computations:
\begin{widetext}
\begin{align}
   &  \mathcal{H}^{I=0,1}(y,\zeta,t,n^2) = \frac{1}{2}\Big[I(y,1,m_\pi^2) + (1-\zeta) I(y-\zeta,1-\zeta,m_\pi^2,n^2)\mp I(y-\zeta,-1,m_\pi^2) \mp (1-\zeta) I(y,\zeta-1,m_\pi^2,n^2)\nonumber\\
    & \;\;\;\; -[(\zeta-2y)m_\pi^2 + t(y-1)] J\!\left(y,\zeta,1,t,m_\pi^2,-\frac{t}{2},n^2\right)\mp[(\zeta-2y)m_\pi^2 + t(y+1-\zeta)] \nonumber
    J\!\left(\zeta-y,\zeta,1,t,m_\pi^2,-\frac{t}{2},n^2\right)\Big]\\
   & \mathcal{E}^{I=0,1}(y,\zeta,t,n^2) = \frac{M}{2}\Big[J\!\left(y,\zeta,1,t,m_\pi^2,-\frac{t}{2},n^2\right)\mp J\!\left(\zeta-y,\zeta,1,t,m_\pi^2,-\frac{t}{2},n^2\right)\Big].\label{tgpdeq}
\end{align}
\end{widetext}

The above formulas are generic for the evaluation based on the diagrams of Fig.~\ref{fig:feyndia}, with the model details (such as the choice of regularization in the NJL model or the selection of parameters) contained in the basic one-loop functions $I$ and $J$. We note the Lorentz invariance and proper crossing symmetry properties. 

In the GPD or tGPD case ($n^2=0$), the above distributions for $I=0$ and $I=1$ differ by functions whose support vanishes in the positive DGLAP region (cf.~Appendix~\ref{app:loops}. Therefore, in the NJL model at the quark-model scale the $I=0,1$ GPDs or tGPDs are equal in the positive DGLAP region, and equal and opposite in the negative DGLAP region. In the symmetric notation
\begin{eqnarray}
&& F^{I=0}(X,\xi,t,n^2=0)={\rm sgn}(X) F^{I=1}(X,\xi,t,n^2=0), \nonumber \\
&& \hspace{5.5cm} {\rm for~} |X|> \xi. \label{eq:eq}
\end{eqnarray}
For $n^2<0$ the formula does not hold, as the supports of the one-loop functions are not separated. Also, the QCD evolution, working differently in the singlet and non-singlet cases, breaks it, so Eq.~(\ref{eq:eq}) holds specifically at the quark-model scale.

In the spontaneously broken phase, the model has three
parameters: the constituent quark mass $M$, the current quark mass
$m$, and the cut-off parameter $\Lambda$ used in the PV regularization
of the scalar loop diagrams. Following the standard procedure,
two of these ($m$ and $\Lambda$) are fixed by demanding particular
values of $m_\pi$ and $f_\pi$, whereas $M$ is set to 300~MeV. The
values of the parameters used in the present work are listed in
Table \ref{partab}. ``Chiral" corresponds to the chiral limit,
``physical" to the charged pion mass, and ``lattice" to a large pion
mass, of the order of the values used in some less expensive lattice
QCD simulations.

\begin{table}[tb]
\caption{The values of the parameters used in the present work and the resulting values of the pion mass, pion weak decay constant, and the pion-quark coupling constant. \label{partab}}
    \centering
    \begin{tabular}{|c|ccc|ccc|}
    \hline
                      & $M$    & $m$     & $\Lambda$ & $m_\pi$  & $f_\pi$  & $g_{\pi q}$ \\ 
                      & [MeV]  & [MeV]  & [MeV]          & [MeV]     & [MeV]    &                  \\ \hline
        chiral     & 300             & 0                &  $731$               &  $0$                  & $86$                 & 3.49 \\
        physical & 300             & 7.5            &  $830$                & $139.6$           &  $93$                 & 3.14 \\
        lattice    & 300             & 41.5              & $1115$               & $400$              & $110$                & 2.29 \\ \hline
    \end{tabular}
\end{table}

\subsection{Quasi GPDs and tGPDs}

In this subsection we present qGPDs and qtGPD obtained in the NJL
model. Since the prime objective of the paper is an investigation of
the dependence of the results on $P_z$, with the lattice feasibility
in mind, we begin by showing in Fig.~\ref{fig:allPz} $H^{I=0,1}$ and
$E^{I=0,1}$ as functions of $Y$. We use here the physical pion mass,
fixed sample values of $\xi=1/2$ and $t=0$, and several representative values of $P_z$. We remark that $P_z \sim 3$~GeV is
about the upper limit accessible in the present-day lattice
studies. Clearly, the limit $P_z\to\infty$ corresponds to GPDs or
tGPDs.  In the model, the GPDs and tGPD have sharp edges at $x=\xi$
and at the end-points $x=\pm
1$~\cite{Broniowski:2007si,Dorokhov:2011ew} (see
Appendix~\ref{app:lim}). Finite value of $P_z$ washes out these
discontinuities, thus covering up the finer details of the
distributions. We note that at $P_z=1$~GeV we are far from the
$P_z\to\infty$ limit, whereas $P_z=3$~GeV gets significantly
closer. The discrepancy is larger for the case of $H^{I=0,1}$ rather
than for $E^{I=0,1}$ .

For finite $P_z$, at asymptotic $Y$ we find for the NJL model with two PV subtractions the behavior
\begin{eqnarray}
&& H^{I=1}\sim |Y|^{-5}, \;\; H^{I=0}\sim Y^{-6}, \nonumber \\
&& E^{I=1}\sim |Y|^{-7}, \;\;E^{I=0}\sim Y^{-8}. \label{eq:asympt}
\end{eqnarray}
According to the discussion in subsection~\ref{sec:polyno}, this limits the ranks of the nonzero $Y$-moments up to 2, 3, 4 and 5  for $H^{I=1}$, $H^{I=0}$, $E^{I=1}$, and $E^{I=0}$, respectively. 

In Fig.~\ref{fig:allmpi} we show an analogous study of the dependence on the pion mass, with fixed $P_z \sim 3$~GeV, and with $\xi=1/2$ and $t=0$. In some lattice QCD simulations one uses a large pion mass, $\sim 300-400$~MeV, which requires less statistics. Thus we present as well the case of a large pion mass, $m_\pi=400$~MeV. The effect of large $m_\pi$ is particularly significant for $E^{I=1}$, but noticeable also for all the other cases.

The effect of changing $\xi$ is not shown separately to not proliferate the number of figures. It just moves the discontinuities of the GPDs or tGPDs at the value $X=\xi$, with the quasi distribution following them, similarly as in Fig.~\ref{fig:allPz}. Naturally, it affects the slopes. The dependence on the momentum transfer $t$ is discussed in detail on the case of the generalized quasi form factors in Sec.~\ref{sec:gff}.

\subsection{Generalized quasi form factors \label{sec:gff}} 

According to Eqs.~(\ref{eq:pol}), qGPDs and qtGPDs may be thought of as infinite collections of the generalized quasi form factors. In this subsection we analyze the lowest generalized quasi form factors, as they should possibly be accessible to lattice determinations. The properties of the generalized form factors related to GPDs, in particular their QCD evolution, were discussed in detail in~\cite{Broniowski:2009zh}, whereas the chiral quark model predictions were reported in~\cite{Broniowski:2008hx}.

\begin{figure}[tb]
    \centering
    \includegraphics[width=0.43\textwidth]{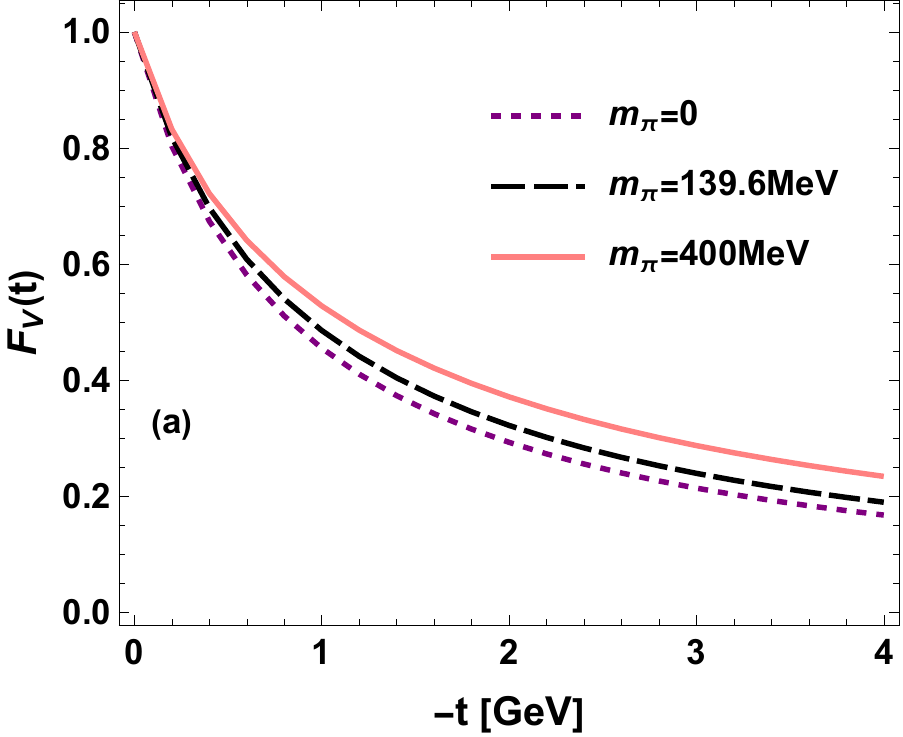}\\~\\
    \includegraphics[width=0.43\textwidth]{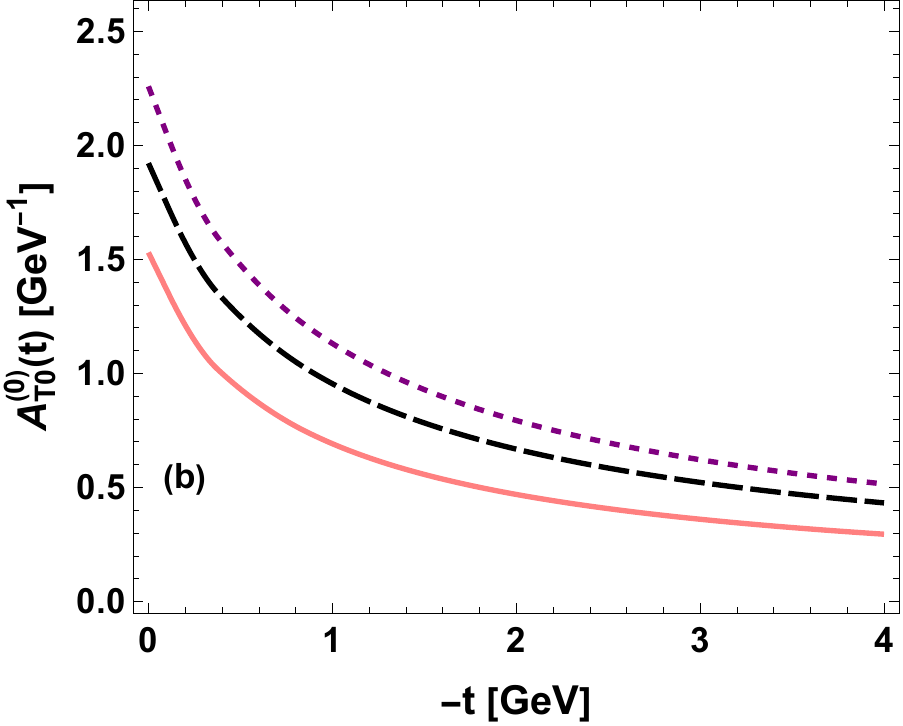}\\~\\
    \includegraphics[width=0.43\textwidth]{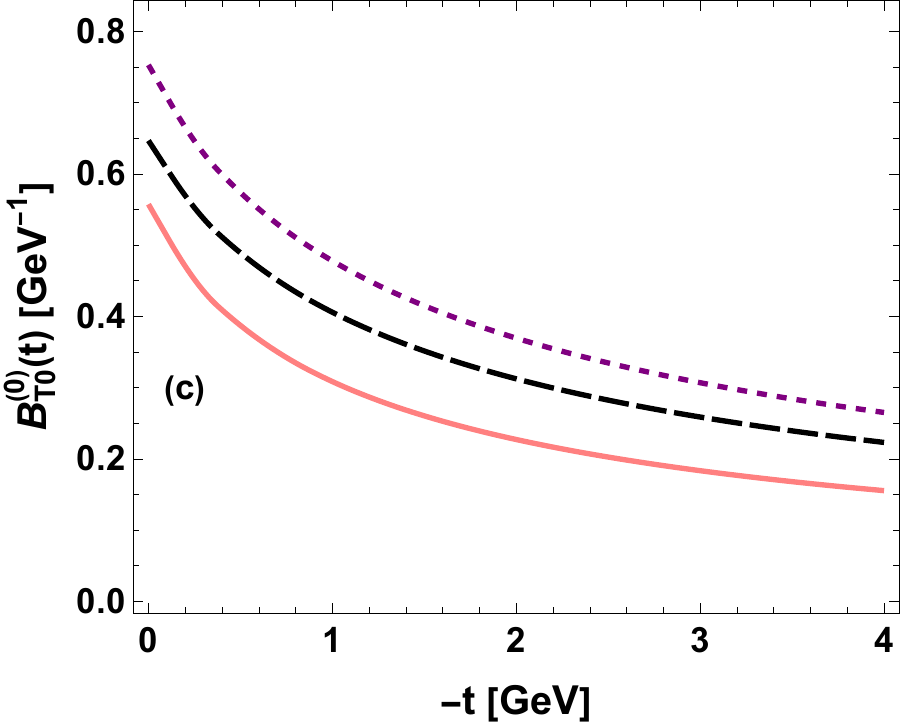}
    \caption{Form factors independent of $n^2$, plotted at space-like momentum transfer $t$ for several values of the pion mass. \label{fig:gff}}
\end{figure}

\begin{table}[tb]
\caption{Values of the rms radii of various form factors for the three considered values of $m_\pi$. \label{tab:rms}}
    \centering
    \begin{tabular}{|l|ccc|}
    \hline
        $m_\pi$ [MeV]                                                                                 & $0$                & $139.6$       &  $400$   \\  \hline 
        $\langle r^2 \rangle_V^{1/2}$   [fm]                                                 & 0.54               & 0.52             &  0.50      \\  
        $\langle r^2 \rangle_{A^{(0)}_{T0}}^{1/2}$   [fm]                              & 0.53               & 0.53             &  0.58      \\
        $\langle r^2 \rangle_{B^{(0)}_{T0}}^{1/2}$   [fm]                              & 0.41               & 0.41             &  0.49      \\   \hline
        $\langle r^2 \rangle_{\theta_1}^{1/2}$   [fm]   $(P_z=1)$                 & 0.57               & 0.56             &  0.62      \\ 
        $\langle r^2 \rangle_{\theta_1}^{1/2}$   [fm]   $(P_z\to \infty)$        & 0.38               & 0.36             &  0.32      \\   \hline
        $\langle r^2 \rangle_{\theta_2}^{1/2}$   [fm]   $(P_z=1)$                 & 0.30               & 0.28             &  0.32      \\  
        $\langle r^2 \rangle_{\theta_2}^{1/2}$   [fm]   $(P_z\to \infty)$        & 0.38               & 0.37             &  0.39      \\    \hline
\end{tabular}
\end{table}

We start with the form factors which are independent of $n^2$, namely $A^{(0)}_0(t)=2F_V(t)$, $A^{(0)}_{T0}(t)$, and $B^{(0)}_{T0}(t)$. They are shown in Fig.~\ref{fig:gff} for three values of $m_\pi$. The mean squared radii are defined generically as 
\begin{eqnarray}
\langle r^2 \rangle = \frac{6}{F(0)} \left . \frac{dF(t)}{dt} \right |_{t=0},
\end{eqnarray} 
and the corresponding root mean squared (rms) radii  are collected for several form factors in Table~\ref{tab:rms}. 

For the vector form factor of Fig.~\ref{fig:gff}(a) we note a mild decrement of the rms radius with $m_\pi$. 
The lower number compared to the experiment, $(0.659(4)~{\rm fm})^2$~\cite{Workman:2022ynf}, is attributed to the missing chiral loops in our leading $1/N_c$ treatment.

\begin{figure}[tb]
    \centering
    \includegraphics[width=0.43\textwidth]{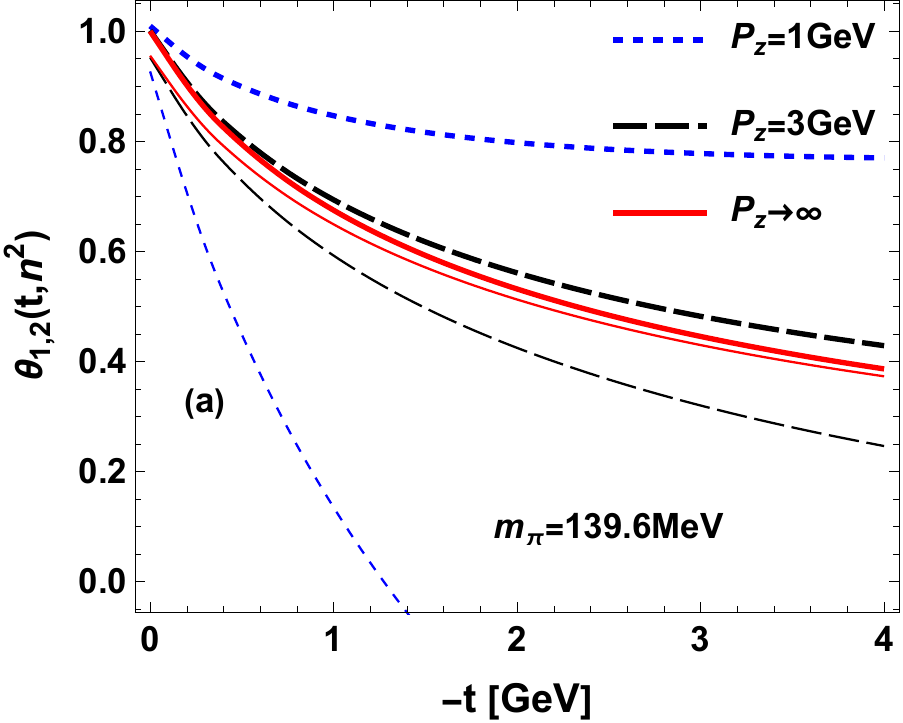}\\~\\
    \includegraphics[width=0.43\textwidth]{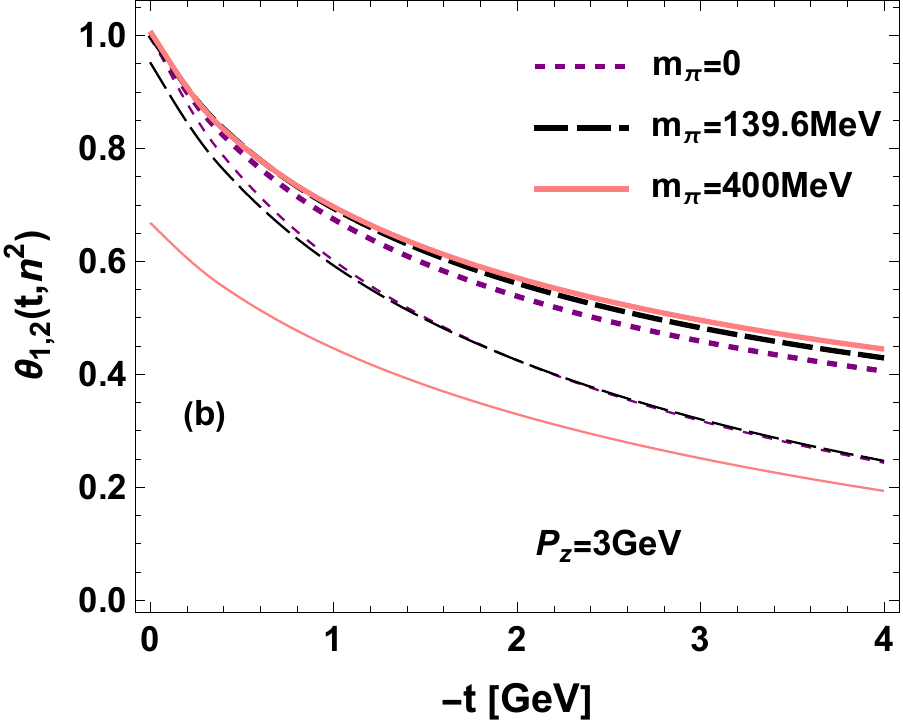}
    \caption{Quasi gravitational form factors $\theta_1$ (thin lines) and $\theta_2$ (thick lines) for (a) various values of $P_z$ at the physical pion mass, and  (b) for various pion masses at $P_z=3$~GeV. \label{fig:grav}}
\end{figure}

\begin{figure}[tb]
    \centering
    \includegraphics[width=0.43\textwidth]{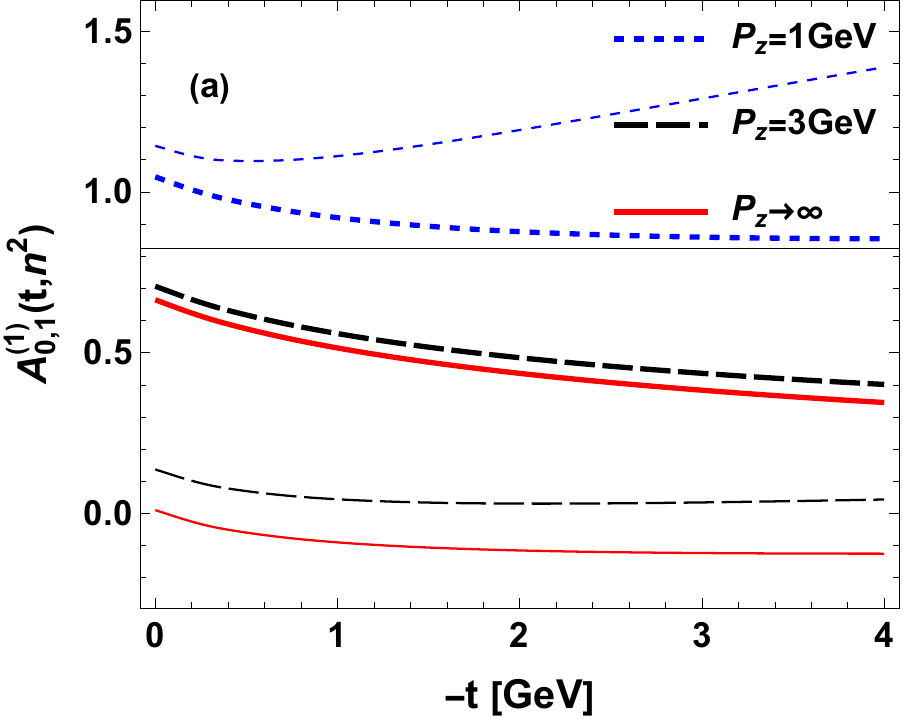}\\~\\
    \includegraphics[width=0.43\textwidth]{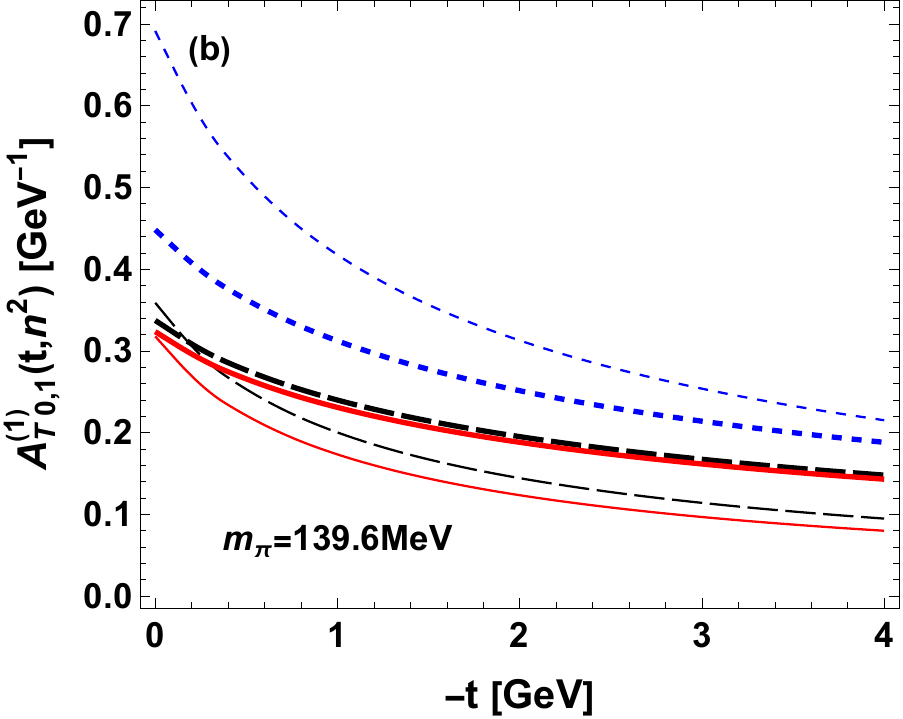}
    \caption{Sample higher-order generalized quasi form factors. \label{fig:high}}
\end{figure}

The lowest order form factors of qtGPDs are shown in Fig.~\ref{fig:gff}(b,c). In our model, the values at the origin are ${\cal N}$ for  $A^{(0)}_{T0}(t)$ and ${\cal N}/3$  for $B^{(0)}_{T0}(t)$, where ${\cal N}$ is given in Eq.~(\ref{eq:normN}). The corresponding rms radii for the $I=1$ form factor, $A^{(0)}_{T0}$, are similar to the vector case.

\begin{figure*}[tb]
\includegraphics[width=0.43\textwidth]{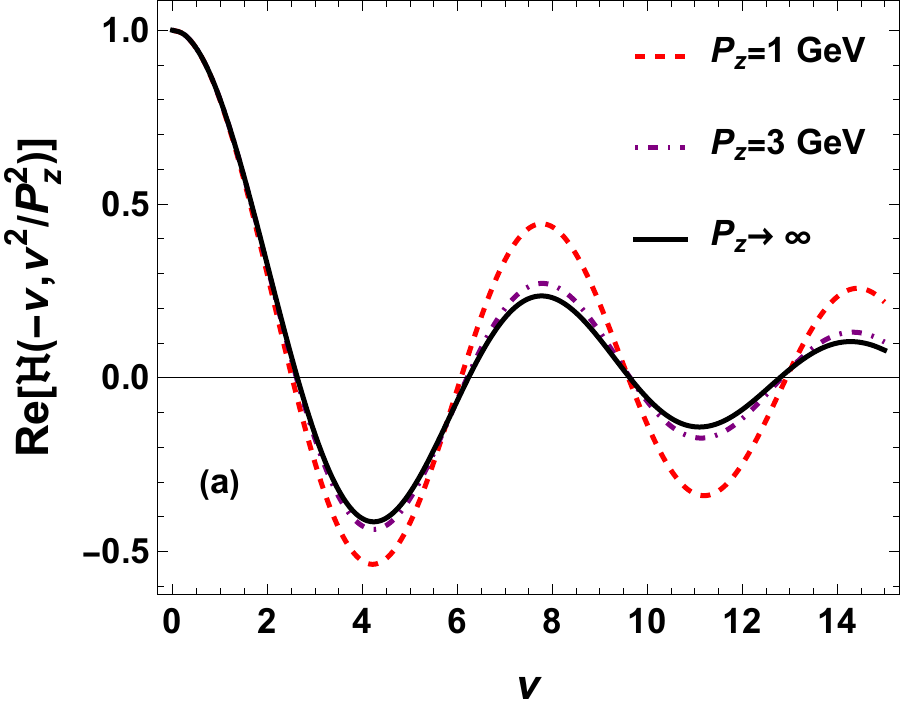} \hfill \includegraphics[width=0.43\textwidth]{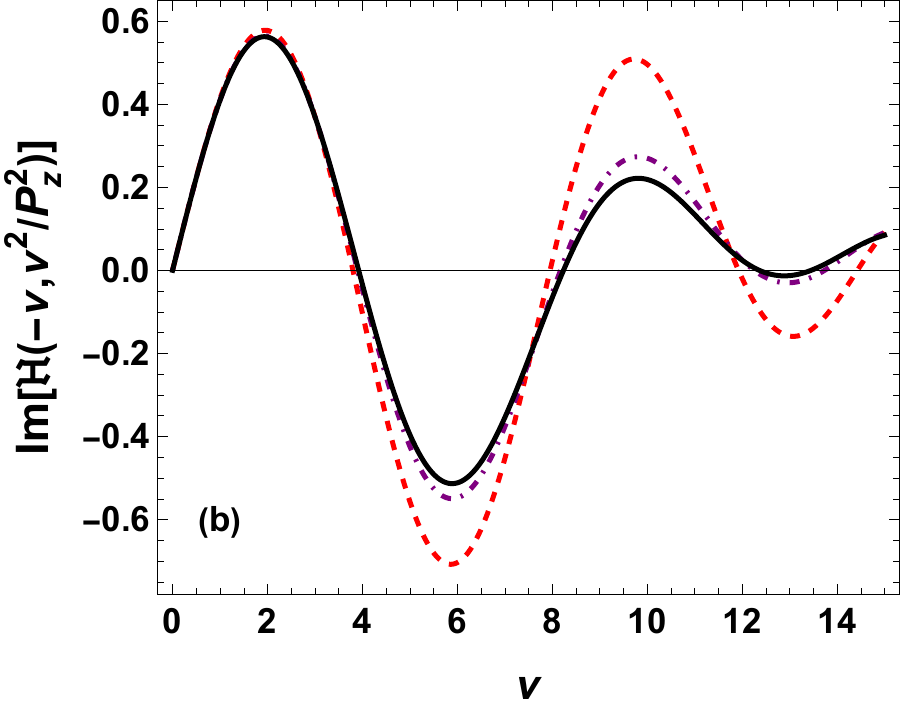} \\ ~ \\
\includegraphics[width=0.43\textwidth]{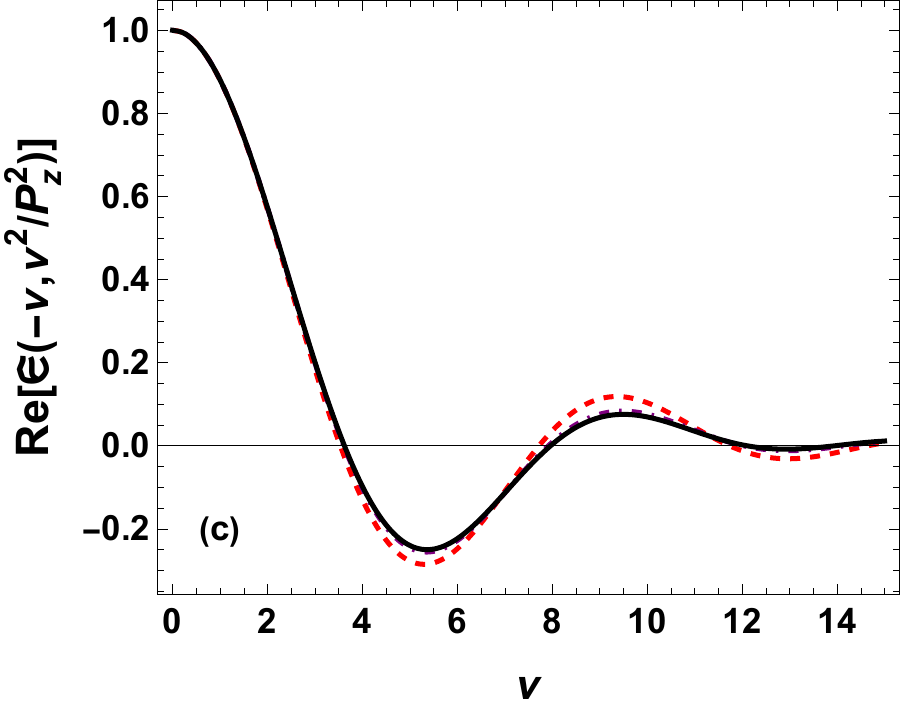} \hfill \includegraphics[width=0.43\textwidth]{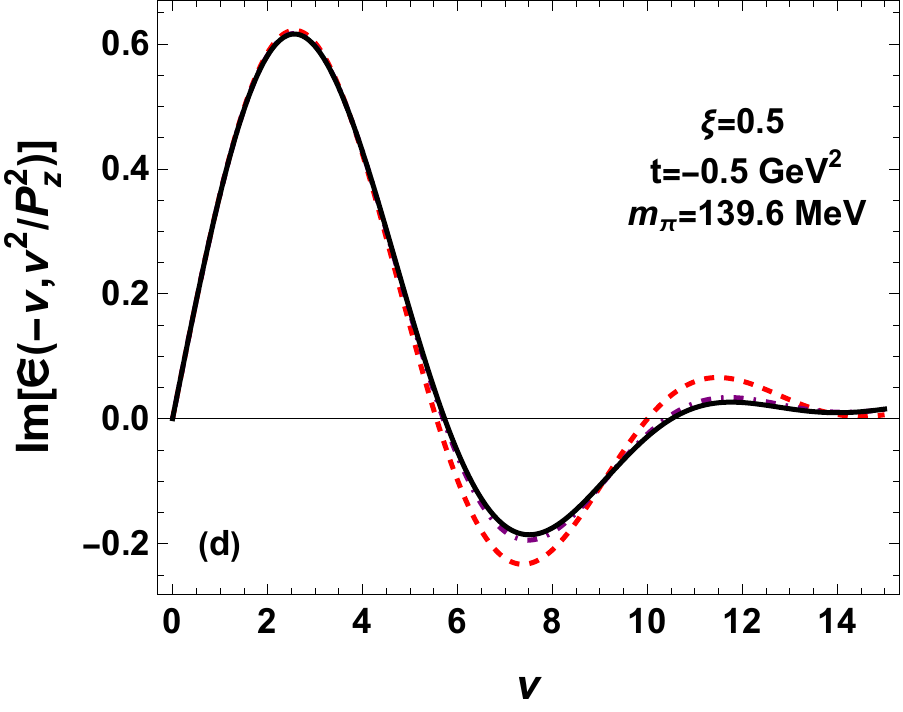} 
   \caption{Real and imaginary parts of the reduced generalized ITDs of the pion, plotted as functions of $\nu$  at $z_3=\nu/P_z$ for various values $P_z$. Physical pion mass, $t=-0.5{\rm GeV}^2$, and $\xi=1/2$. \label{fig:ITDpz}}
\end{figure*}

The two quasi gravitational form factors are presented in Fig.~\ref{fig:grav}. Unlike the above-discussed cases, these form factors do depend on $P_z$. What is independent of $P_z$, however, is the value of $\theta_2$ at the origin,
$\theta_2(0,n^2)=1$. This model result is more general than the energy momentum sum-rule mentioned in subsection~\ref{sec:polyno}, holding for the $n^2=0$ case. Also, in our model in the chiral limit $\theta_1(t,n^2)-\theta_2(t,n^2) = {\cal O}(m_\pi^2)$ for any $n^2$, extending the similar low energy theorem for light-like vectors, $n^2 =0$.

The values of the rms radii of the quasi gravitational form factors given in Table~\ref{tab:rms} exhibit a strong dependence on $P_z$, which is complementary to the behavior in Fig.~\ref{fig:grav}. We note that the model relation showing a more compact distribution of matter than charge in the pion~\cite{Broniowski:2008hx},
\begin{eqnarray}
2\langle r^2 \rangle_{\theta_{1,2}}=\langle r^2 \rangle_V,
\end{eqnarray}
holds for $n^2=0$ and in the chiral limit. 

There is yet another relevant feature pertaining to the form factors
discussed up to now. The vector form factor, corresponding to a
conserved current, does not evolve with the QCD scale. The QCD
evolution of the generalized form factors (for $n^2=0$) is
multiplicative for the gravitational form factors (albeit the quark pieces
mix with the gluon distributions), hence the valence form factors
change only by an overall factor, whereas their shape in $t$ is
preserved. Analogously, for the rank-0 form factors of $E^{I=0,1}$ the
renormalization is multiplicative, hence their shape in $t$ is
preserved. This feature is generally not preserved for the higher rank
generalized form factors, whose LO QCD evolution mixes
them~\cite{Broniowski:2009zh}. Therefore we encounter a situation
where the higher rank generalized quasi form factors not only strongly
depend on $n$, but also are largely affected by the QCD evolution. For
their assessment at higher scales, carrying out the evolution is necessary.

\subsection{Generalized Ioffe-time distributions \label{sec:itd}}

\begin{figure*}[tb]
\includegraphics[width=0.43\textwidth]{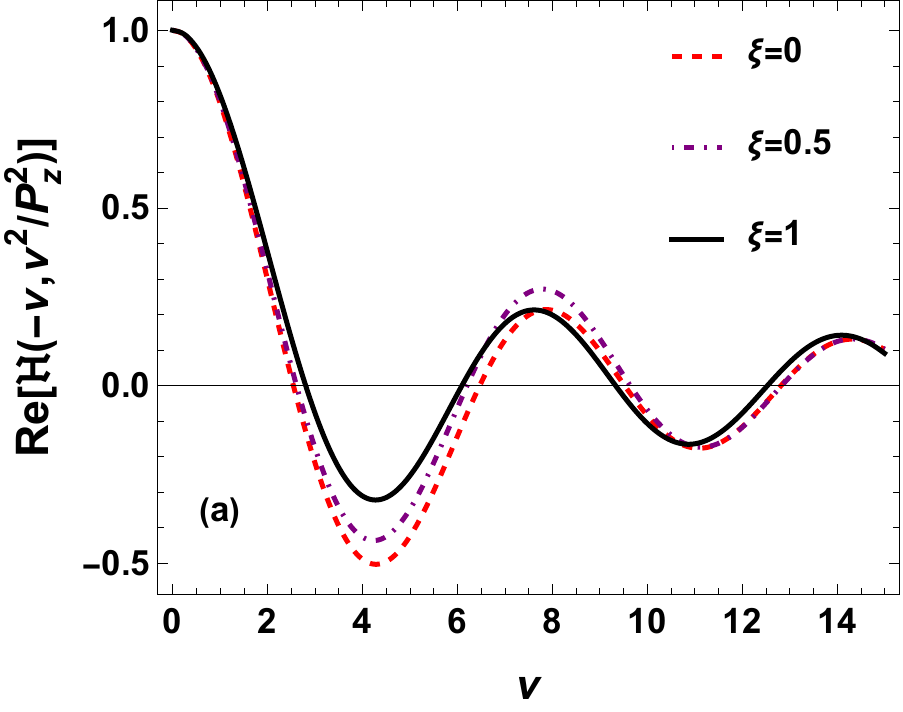} \hfill \includegraphics[width=0.43\textwidth]{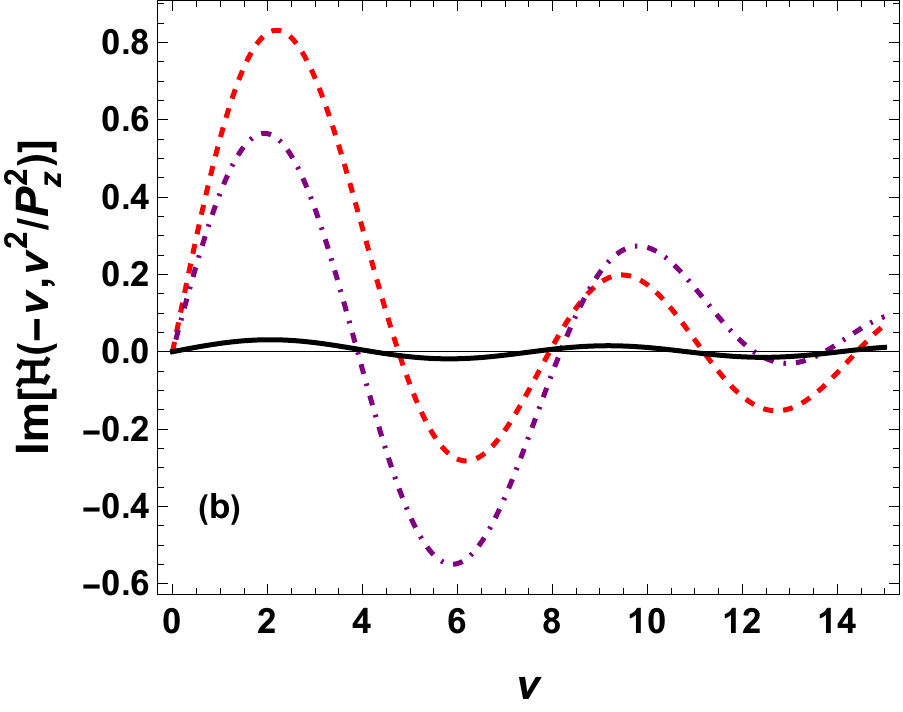} \\ ~ \\
\includegraphics[width=0.43\textwidth]{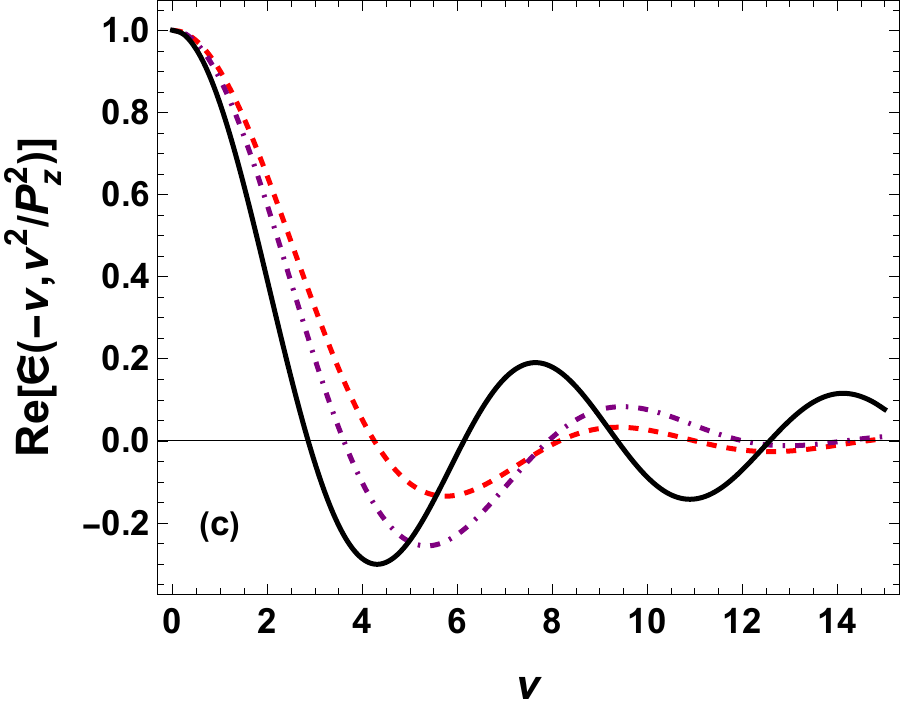} \hfill \includegraphics[width=0.43\textwidth]{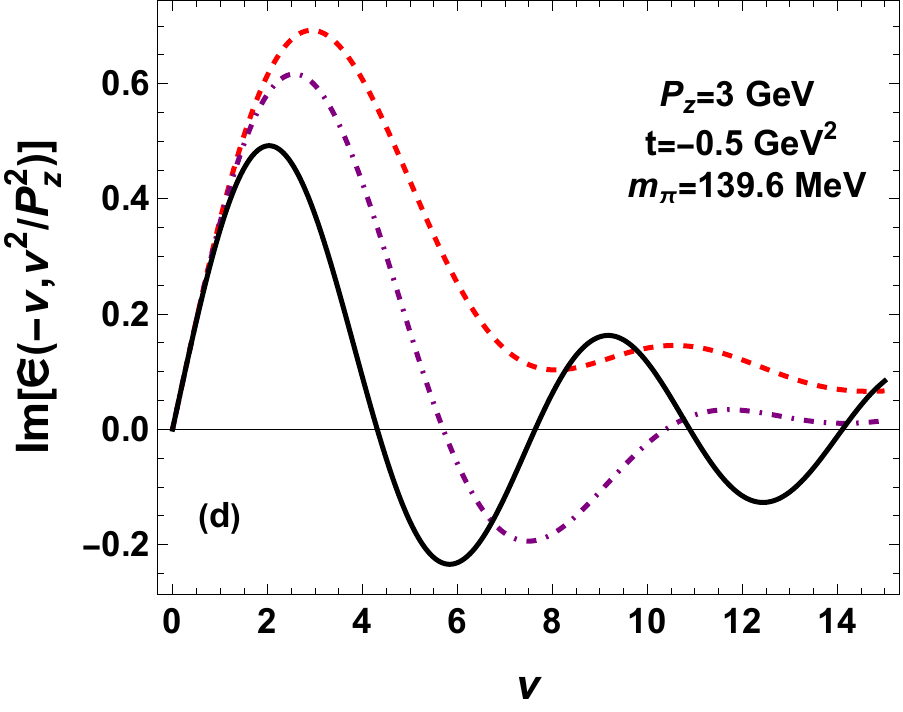} 
   \caption{Same as in Fig.~\ref{fig:ITDpz} but for different values of the skewness parameter $\xi$. Physical pion mass, $t=-0.5~{\rm GeV}^2$, and $P_z=3$~GeV. \label{fig:ITDxi}}
\end{figure*}

\begin{figure*}[tb]
\includegraphics[width=0.43\textwidth]{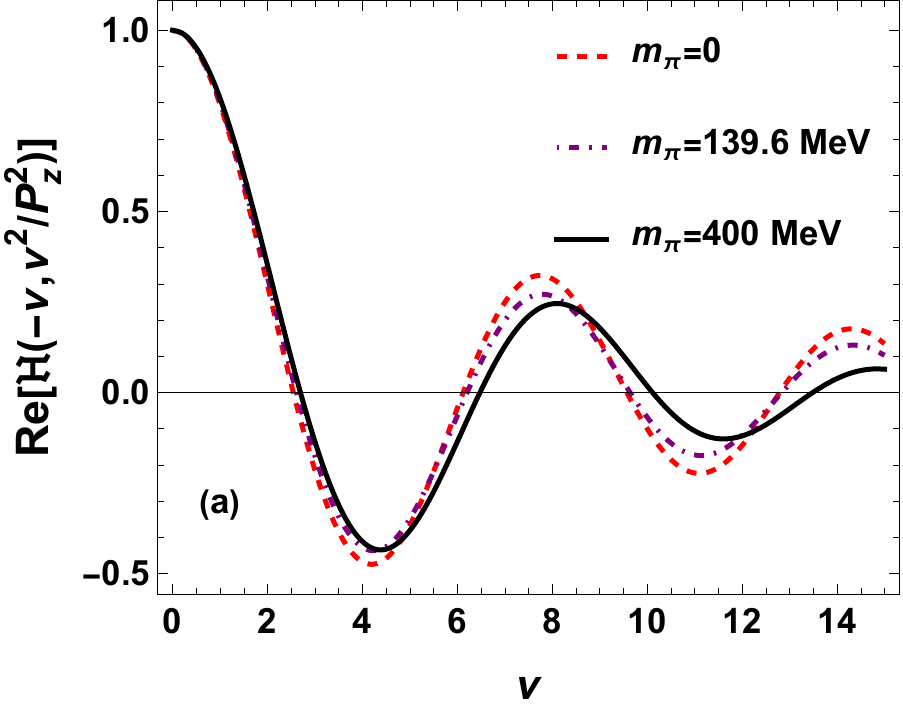} \hfill \includegraphics[width=0.43\textwidth]{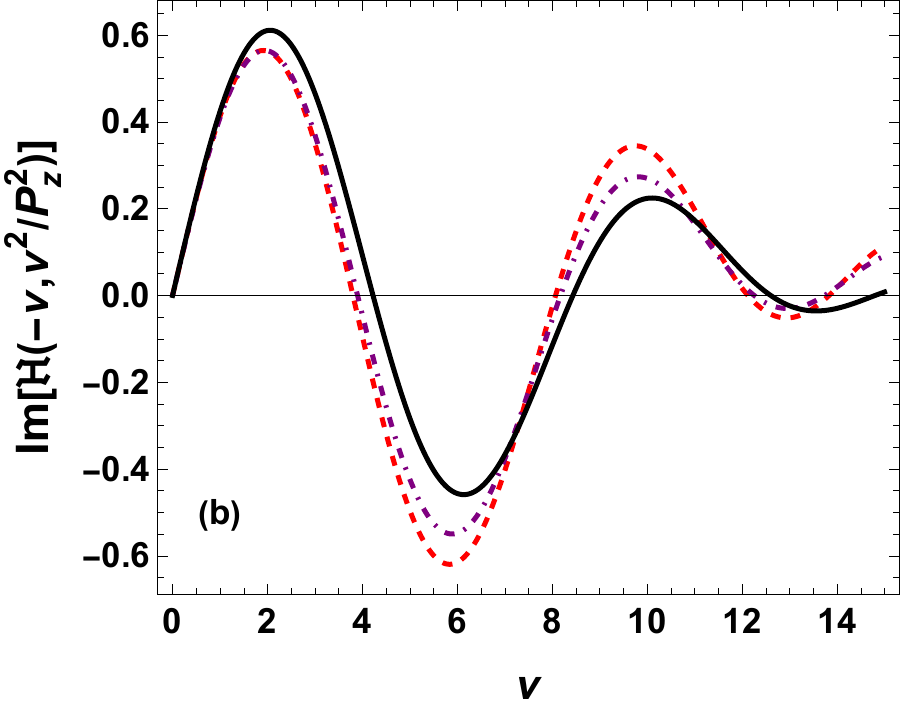} \\ ~ \\
\includegraphics[width=0.43\textwidth]{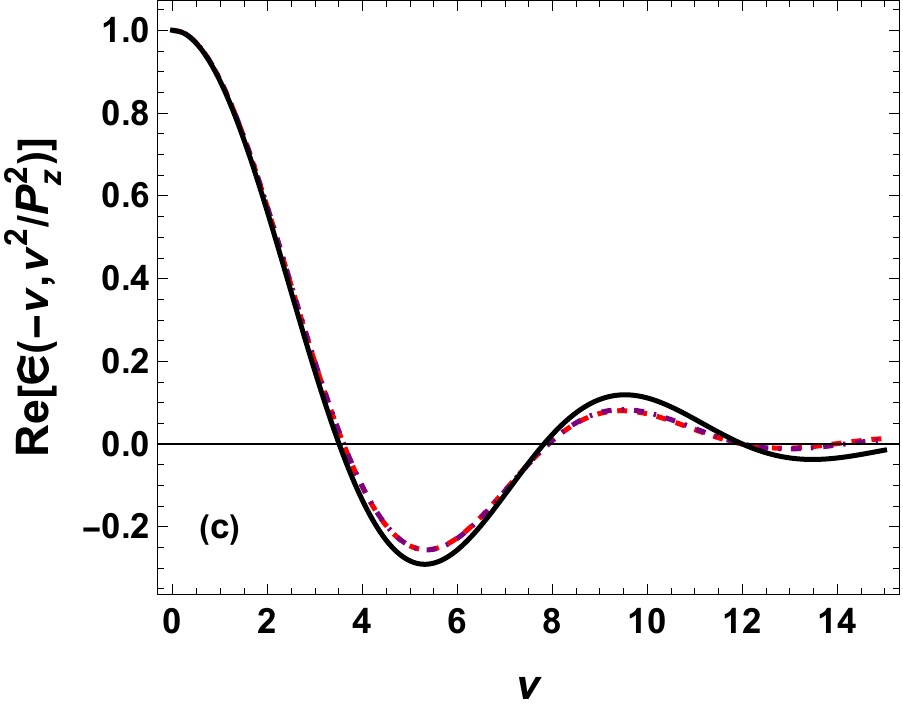} \hfill \includegraphics[width=0.43\textwidth]{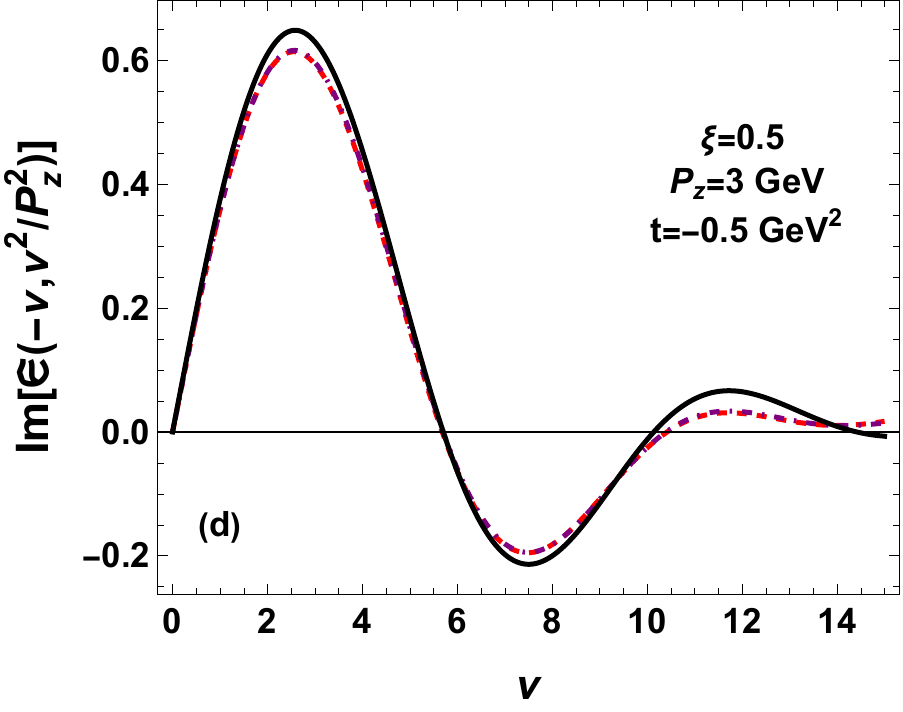} 
   \caption{Same as in Fig.~\ref{fig:ITDpz} but for different values of the pion mass at $\xi=0.5$, $t=-0.5~{\rm GeV}^2$, and $P_z=3$~GeV. \label{fig:ITDmpi}}
\end{figure*}

\begin{figure*}[tb]
\includegraphics[width=0.43\textwidth]{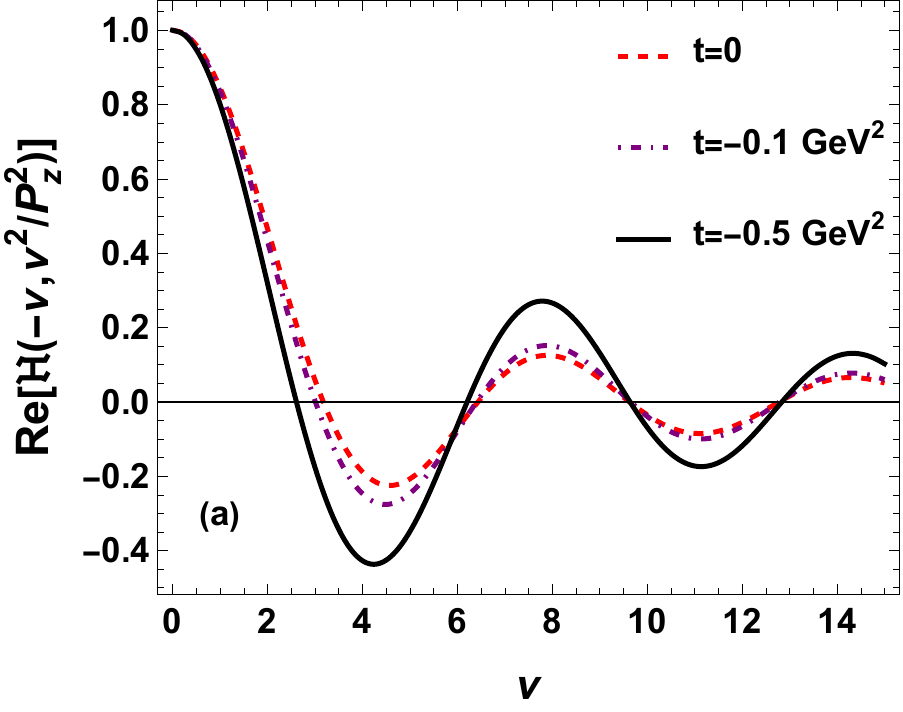} \hfill \includegraphics[width=0.43\textwidth]{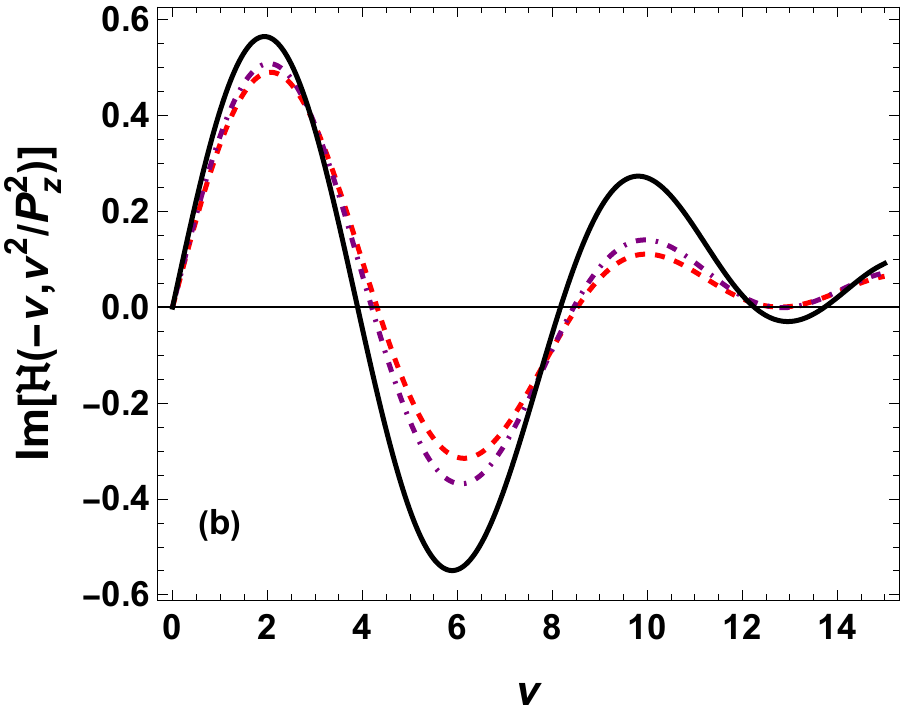} \\ ~ \\
\includegraphics[width=0.43\textwidth]{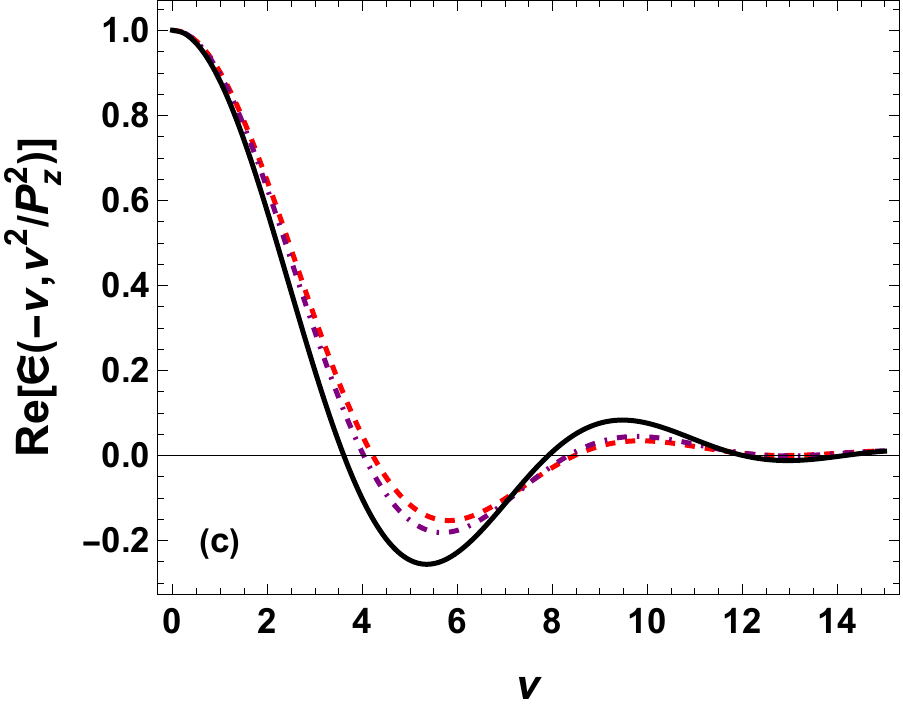} \hfill \includegraphics[width=0.43\textwidth]{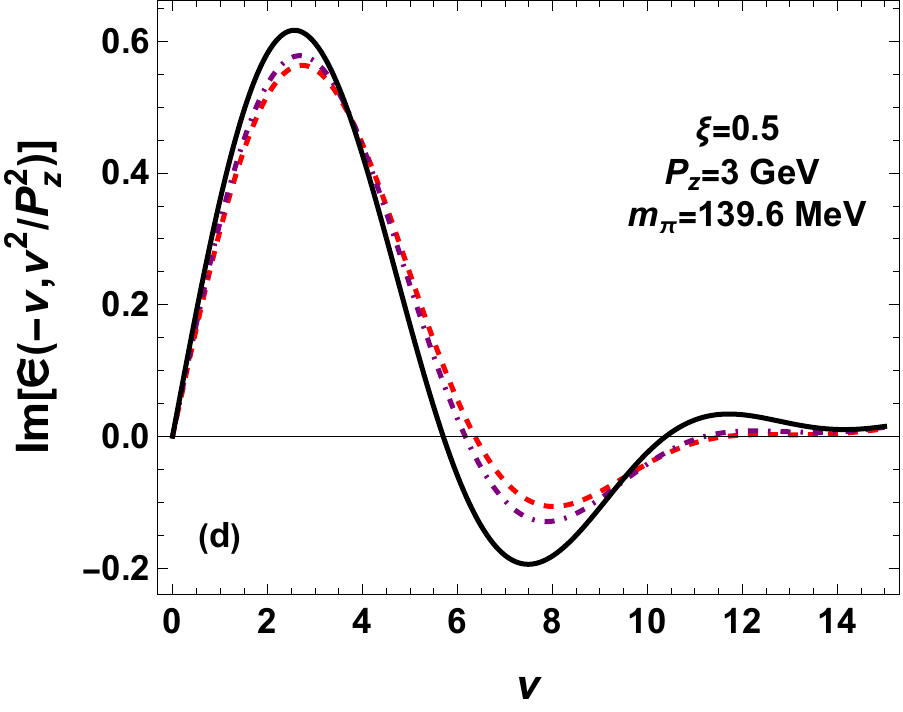} 
   \caption{Same as in Fig.~\ref{fig:ITDpz} but for different momentum transfer $t$ at the physical pion mass, $\xi=0.5$, and $P_z=3$~GeV. \label{fig:ITDt}}
\end{figure*}

In this subsection we show our model results for the reduced generalized ITDs defined in Eq.~(\ref{eq:redI}) for each of the functions $H^{I=1,0}$ and  $E^{I=1,0}$. We discuss them in detail, as they have become basic objects of the lattice QCD studies.
From symmetry properties in $\nu$, the $I=1$ parts are purely real, and the $I=0$ parts purely imaginary, so we do not have to carry the isospin labels in the notation. The (reduced) ITD of Eq.~(\ref{eq:redI}) are in general functions of, independently, $-\nu$ and $-z^2$, but following~\cite{Orginos:2017kos} we take the sections defined by $-z^2=\nu^2/P_z^2$. First, we investigate the dependence of the reduced ITDs on $P_z$.

In the NJL model, the one-loop expressions for the generalized ITDs follow from the simple formulas of Appendix~\ref{app:ioffe}. It is clear that in the chiral limit and for $t=0$ the dependence on $z_3=\sqrt{-z^2}$ and $\nu$ is separated, as $z_3$ appears only in the argument of the Bessel functions $K_{0,1}(z_3 M)$. Therefore the dependence on $z_3=\nu/P_z$ cancels out from the ratio in Eq.~(\ref{eq:redI}) and the results for $\mathfrak{H}$ and $\mathfrak{E}$ do not depend on $P_z$. On the contrary, the ITDs (not reduced) would display a strong non-separability of $\nu$ and $z_3$, obscuring the picture~\cite{Orginos:2017kos}.

In the chiral limit and for $t=0$, the following formulas are derived in Appendix~\ref{app:lim}:
\begin{eqnarray}
&& \mathfrak{H}= \frac{\sin \nu}{\nu} + i \frac{\cos \nu \xi - \cos \nu}{\nu}, \nonumber \\
&& \mathfrak{E}= 2\frac{\cos \nu \xi - \cos \nu}{\nu^2(1-\xi^2)} + 2 i \frac{\xi \sin \nu - \sin \nu \xi}{\nu^2 \xi (1-\xi^2)}, \label{eq:iofgpd}
\end{eqnarray}
where the real and imaginary parts correspond to the $I=1$ and $I=0$ pieces, respectively. Of course, the independence of $P_z$ here is manifest.

Nonzero values of $m_\pi$ or $t$ cause a due breaking of the $\nu$--$z_3$ separability, as can be clearly seen from Fig.~\ref{fig:ITDpz} made for the physical pion mass, a moderate $t=-0.5~{\rm GeV}^2$, and $\xi=1/2$. The breaking is more prominent at larger values of $\nu$, as expected from the fact that the second argument is $\nu^2/P_z^2$. While for the displayed range of $\nu$ the difference between $P_z=1$~GeV and  $P_z \to \infty$ is substantial, the results for $P_z=3$~GeV essentially coincide with the GPD limit of $P_z \to \infty$.

Next, we pass to studying the dependence of the generalized ITDs on $\xi$. The expansion of formulas~(\ref{eq:iofgpd}) at $\nu=0$ yields (at $m_\pi=0$ and $t=0$)
\begin{eqnarray}
&& \mathfrak{H}= \sum_{k=0}^\infty \frac{\nu^{2k}}{(2k+1)!} + i \sum_{k=0}^\infty (-1)^k\frac{\nu^{2k+1}(1-\xi^{2k+2})}{(2k+2)!}, \nonumber \\
&& \mathfrak{E}= \sum_{k=0}^\infty (-1)^k \frac{2\nu^{2k}p_k(\xi^2)}{(2k+2)!} + i \sum_{k=0}^\infty (-1)^k\frac{2\nu^{2k+1}p_k(\xi^2)}{(2k+3)!}, \nonumber \\
\label{eq:iofexpexa}
\end{eqnarray}
where $p_k(\xi^2)=1+\xi^2+\dots+\xi^{2k}$, and explicitly
\begin{eqnarray}
&& \mathfrak{H}= 1-\tfrac{1}{6}\nu^2  + i \left [ \tfrac{1}{2} (1-\xi^2) \nu - \tfrac{1}{24}(1-\xi^4) \nu^3\right ]+..., \nonumber \\
&& \mathfrak{E}= 1-\tfrac{1}{12}(1+\xi^2) \nu^2 + i \left [ \tfrac{1}{3} \nu - \tfrac{1}{60}(1+\xi^2) \nu^3 \right ]+.... \nonumber \\
\label{eq:iofexp}
\end{eqnarray}
We note the manifestation of polynomiality in the above expressions in their dependence on $\xi$. In the case of general $m_\pi$ and $t$, in the limit of $P_z\to\infty$ the coefficients of the expansion of the generalized ITDs relate to the $X$-moments of the corresponding GPDs, hence provide the information on the 
generalized form factors.

For the case shown in Fig.~\ref{fig:ITDxi}, prepared with the physical pion mass, $t=-0.5$, and $P_z=3$~GeV, the results are not too far away from the case of Eqs.~(\ref{eq:iofgpd}), hence we note the characteristic  approximate features, such as the weak dependence of ${\rm Re}(\mathfrak{H})$ on $\xi$, the nearly zero ${\rm Im}(\mathfrak{H})$,  the proportionality of the slope of ${\rm Im}(\mathfrak{H})$ to $(1-\xi^2)$, the increase of the curvature of ${\rm Re}(\mathfrak{E})$ as $(1+\xi^2)$, and the independence of the slope of ${\rm Im}(\mathfrak{E})$ of $\xi$. 
In general, we note that, as expected from the framework, the dependence on $\xi$ is certainly a relevant feature. Methodologically, this can be used as an alternative way to obtain the generalized form factors. 

In Figs.~\ref{fig:ITDmpi} and \ref{fig:ITDt} we examine the dependence of the generalized ITDs on the value of $m_\pi$ and $t$ for the fixed value of $P_z=3$~GeV. The features directly reflect the appearance of $m_\pi$ and $t$ in Eqs.~(\ref{tgpdeq}). We find that the dependence on $m_\pi$ is very moderate, even with the large value of $m_\pi=400$~MeV. The differences are a bit larger for  
$\mathfrak{H}$ than for $\mathfrak{E}$. On the other hand, the dependence on $t$ is somewhat more substantial, partly because we have taken as moderate the value of $-t=0.5~{\rm GeV}^2$, which, however, is significantly larger than the used values of $m_\pi^2$.

\subsection{Generalized pseudo-distributions \label{sec:pseudo}}

\begin{figure}[tb]
\includegraphics[width=0.43\textwidth]{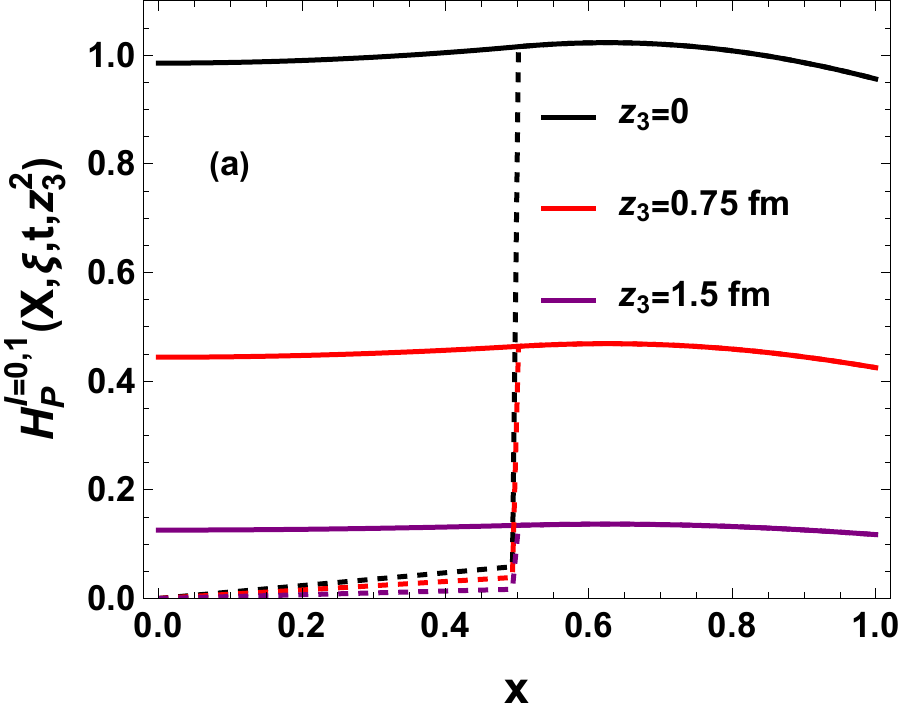} \\
\includegraphics[width=0.43\textwidth]{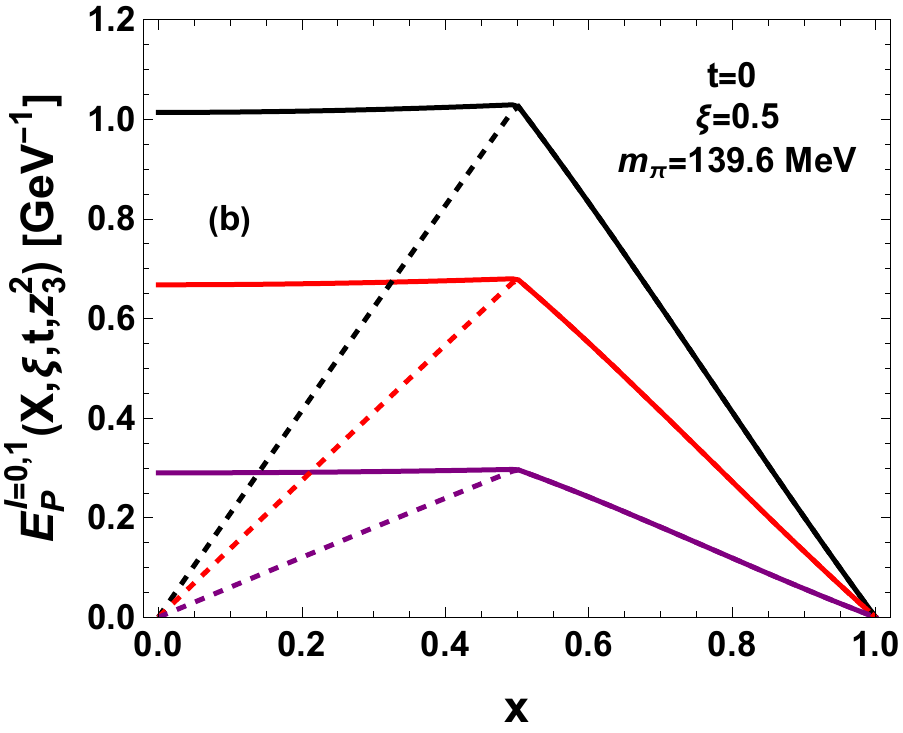}  
   \caption{Generalized pseudo-distributions for the physical pion mass, $t=0$, and $\xi=0.5$, plotted as functions of the momentum fraction $x$ for three sample values of $z_3$. \label{fig:pseudo}}
\end{figure}

\begin{figure}[tb]
\includegraphics[width=0.43\textwidth]{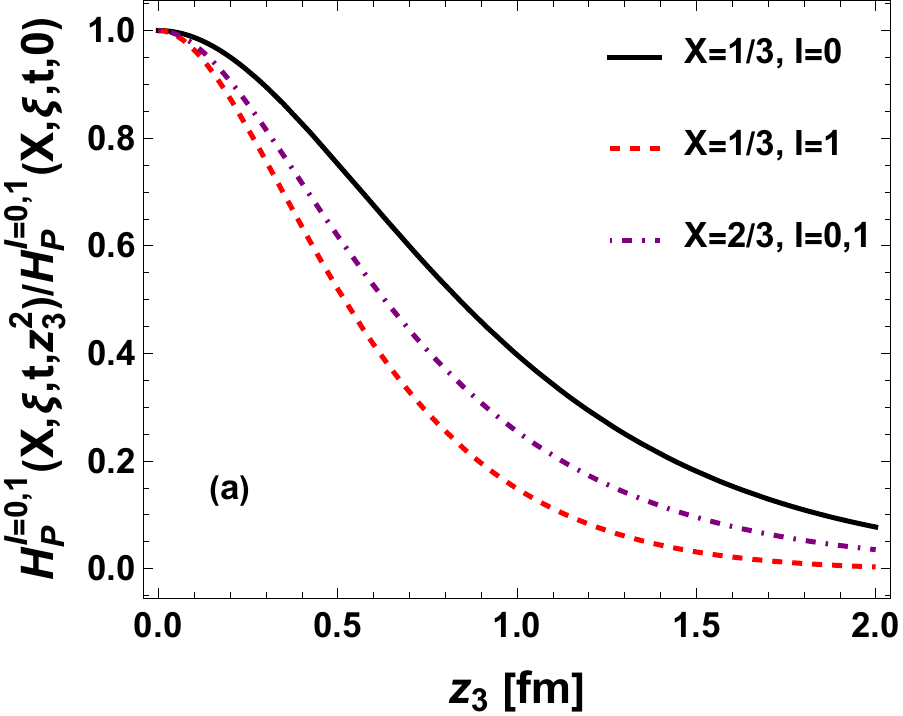}  \\
\includegraphics[width=0.43\textwidth]{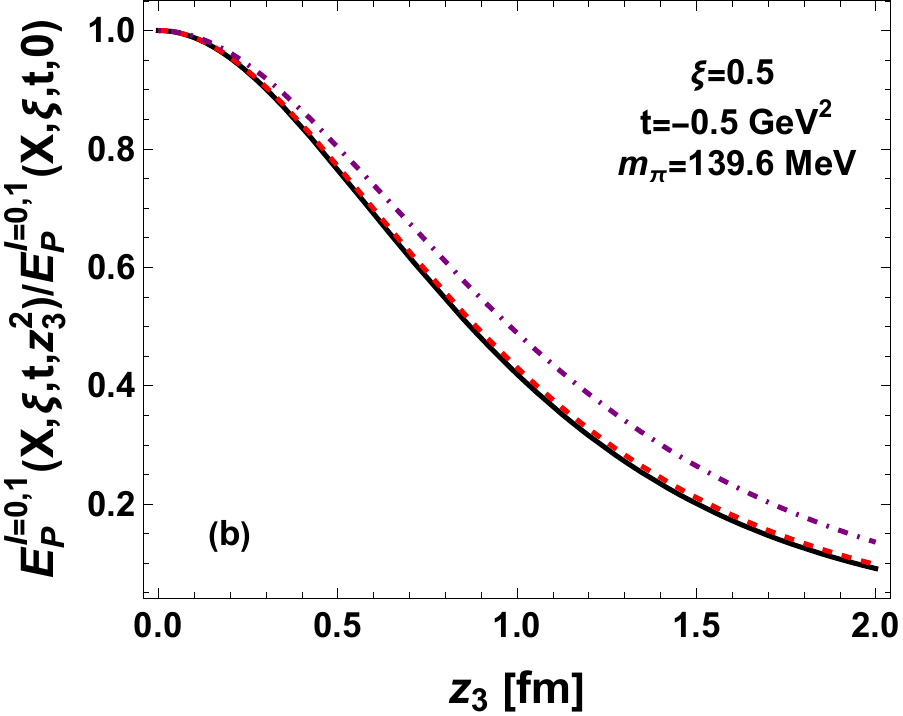}  \\
   \caption{Pseudo form factors $F_P(x,\xi,t,z_3^2)/F_P(x,\xi,t,0)$ for the physical pion mass, $t=0$, and $\xi=0.5$, plotted as functions of $z^3$ for two sample values of $x$: one in the DGLAP and one in the ERBL region. \label{fig:pseudoff}}
\end{figure}

First, we notice that equalities analogous to Eq.~(\ref{eq:eq}) hold for the generalized pseudo-distributions in the DGLAP region for {\rm any} value of $z_3$:
\begin{eqnarray}
&& F_P^{I=0}(X,\xi,t,-z_3^2)={\rm sgn}(X) F_P^{I=1}(X,\xi,t,-z_3^2), \nonumber \\
&& \hspace{5.5cm} {\rm for~} |X|> \xi. \label{eq:eqp}
\end{eqnarray}
The reason is a proper separation of the supports of the loop functions provided in Appendix~\ref{app:ioffe}

In Fig.~\ref{fig:pseudo} we show the generalized pseudo-distributions of Eq.~(\ref{eq:psdef}) obtained in our model for the physical $m_\pi$, $t=0$, and skewness $\xi=1/2$, plotted as functions of $x$ for three sample values of $z_3$. We can vividly see that in the DGLAP region the generalized pseudo-distributions coincide, according to Eq.~(\ref{eq:eqp}). The case of $z_3 \to 0$ corresponds, naturally, to the GPDs or tGPDs. We note that with increasing $z_3$, the distributions decrease, which is naturally attributed to the pseudo form factor in $z_3$, defined as $F_P(x,\xi,t,z_3^2)/F_P(x,\xi,t,0)$.

From the form of Eqs.~(\ref{eq:psI}) and (\ref{eq:psJ}) it is clear
that in our model, for $m_\pi=0$ and $t=0$, the dependence on $x$ and
$z^2=-z_3^2$ factorizes. Then the shape of the generalized
pseudo-distributions is given by Eqs.~(\ref{eq:h1s},\ref{eq:e1s}),
supplied with a universal ($x$-independent) pseudo form factor in
$z_3^2$. For the case of a general kinematics the factorization no
longer holds exactly, but it still nearly holds for the physical value of
$m_\pi$ and small values of $-t$.  At larger values of $-t$ it becomes
visibly broken. This is manifest in Fig.~\ref{fig:pseudoff}, where the
form factor is plotted for the physical pion mass, $\xi=0.5$, and
$t=-0.5~{\rm GeV}^2$, for several values of $x$, both in the DGLAP and
ERBL region. In the DGLAP region (here we take $x=2/3$), the form factors for
the $I=0$ and $I=1$ cases are identical, as follows directly from
Eq.~(\ref{eq:eqp}). In the ERBL region we notice the breaking, which
essentially is due to the taken value of $-t$ (whereas $m_\pi^2\simeq
0.02~{\rm GeV}^2$, a very small number).

We also note from Fig.~\ref{fig:pseudoff} that the breaking is stronger for the $H$ pseudo form factors than for the $E$ case. This feature reflects the structure of Eq.~(\ref{tgpdeq}), where $t$ multiplies the $J$ functions in the definition of $H$, but not for $E$, thus contributing to stronger breaking of the factorization.

As the Fourier transform from ${\bm z}$ to ${\bm k}_T$  converts the generalized pseudo-distributions into the generalized $k_T$-unintegrated distributions (cf. Eq.~(\ref{eq:kT})), the plots of Fig.~\ref{fig:pseudoff} provide also complementary information on the $k_T$-unintegrated GPDs and tGPDs. In particular, the curvature of the pseudo form factors at the origin yields the average transverse momentum squared, $\langle k_T^2 \rangle(x,\xi,t)$, in the $k_T$-unintegrated GPDs and tGPDs.

Finally, we wish to discuss the range of $|\nu|$, denoted as $\nu_{\rm max}$, needed to carry out the inverse
Fourier transform in definition~(\ref{eq:psdef}). This is of practical
importance, as in numerical evaluation such as in the lattice QCD
simulations, unlike our analytic case, we always have an upper bound
for $|\nu|$. From Fig.~\ref{fig:pseudo} we can see that the rate of
variation of the pseudo-distributions is about $\Delta={\rm min}(\xi,
1-\xi)$ for $0<\xi<1$, and $\Delta=1$ for $\xi=0$ or $\xi=1$ (the pseudo-PDF or the pseudo-DA limit). For the Fourier transform to be able to reproduce
it, we need $\nu_{\rm max} \gg 2\pi/\Delta$. Thus, for a fixed
$\nu_{\rm max}$ more accurate results would follow for $\xi=1/2$ than
for $\xi$ close (but not equal) to 0 or 1. For $\xi=1/2$ we need
$\nu_{\rm max} \gg 4\pi$, so at least of the order of 30, depending on
the demanded accuracy. Such large values are presently prohibitive for
the lattice QCD simulations.

We remark that a similar problem occurs in the ``good lattice cross section'' method~\cite{Ma:2014jla,Ma:2017pxb}, where also large values of the Ioffe-time are necessary~\cite{Broniowski:2020had} to provide a reliable information on the higher $x$-moments. 

\section{Conclusions}

We have carried out an extensive analysis of the quasi GPDs and tGPDs, related generalized Ioffe-time distributions, and generalized pseudo distributions of the pion in the framework of the NJL model. Even in this very simple model, treated at the one-quark-loop (large-$N_c$) level, the results are nontrivial and reveal the rich structure of the examined objects. 

Of particular interest is the dependence of the results on the
momentum of the pion, $P_z$. Referring to the ITDs, which are the quantities directly accessible
in the lattice QCD simulations, we find that $P_z=3$~GeV is sufficiently close (within a few percent)
to the desired limit of $P_z\to\infty$ for low-enough values of the Ioffe time $\nu<15$, while
the case $P_z=1$~GeV is still significantly away (cf. Fig.~\ref{fig:ITDpz}). This is especially so for the case of
non-zero skewness, where more variation is present in the
distributions, in particular the isoscalar GPDs. We stress that the conclusion that  $P_z\simeq3$~GeV is large enough for ITDs does not carry over to the qGPDs themseves (cf. Fig.~\ref{fig:allPz}), where much larger values would be necessary, say, $P_z>10$~GeV, depending on how closely one wishes to approximate the  $P_z\to\infty$ limit near the endpoints and near $Y=\pm \xi$. The issue is related to the general difficulties in accurately Fourier transforming the ITDs into qGPDs with a limited accessible range in the Ioffe time, as discussed at the end of Sec.~\ref{sec:pseudo}.

Our estimates may be useful for implementations of the skewed quasi distributions on the
lattice, which are yet to come. We have also studied the dependence on
the pion mass, including a large value of 400~MeV, which is in the
range of values used some lattice QCD studies.

We have computed quasi generalized form factors, related to moments of the qGPDs and qtGPDs, and investigated their dependence on $P_z$. Only the lowest rank form factors are independent of $P_z$, and the higher ones do, in particular, the gravitational form factors, whose shape changes strongly between $P_z=1$~GeV and 3~GeV. They are also sensitive to $m_\pi$.

The generalized (reduced) Ioffe-time distributions, basic objects for the lattice investigations, encode the information on the quasi-distributions as their Fourier transforms. We have discussed in detail the dependence on the skewness parameter $\xi$, exhibiting the simple characteristics such as the slope or curvature of certain GPDs or qGPDs. 

We have also estimated how large values of the Ioffe time $\nu_{\rm max}$ are needed to effectively invert the Fourier transform to get the pseudo-distributions defined in the $x$-space. The value of $\nu_{\rm max}$, should be larger if skewness is present, as it causes larger variation in the qGPDs and qtGPDs. Roughly, values $\nu_{\rm max}>30$ are necessary.

With the obtained pseudo distributions, we have investigated the breaking of the longitudinal-transverse separability. The effects of $m_\pi$ are rather small here, but larger values of the momentum transfer $t$ cause significant breaking. The issue is related to properties of the $k_T$-unintegrated distributions (or TMDs).

We stress that our analysis is based on analytic or semi-analytic expressions, which
allows for a simple insight and illustration of the intricate formal
features of the field.  At the same time, our results pertain to the
quark-model scale, which is much lower from the experimental or
lattice scales. In the context of the QCD evolution, we have thus
investigated the initial conditions and their sensitivity to $P_z$,
$m_\pi$, $t$, or $\xi$. That sensitivity will be carried over to higher
scales, with the exact effects to be estimated in a future study. With
the existing link to the $k_T$-unintegrated distributions, the
QCD evolution of the latter ones can be used to evolve the qGPDs and
qtGPDs.

\begin{acknowledgments}
  VS acknowledges the support by the Polish National Science Centre
  (NCN), grant 2019/33/B/ST2/00613. WB acknowledges the support by the
  Polish National Science Centre, grant 2018/31/B/ST2/01022. ERA
  acknowledges support from project PID2020-114767GB-I00 funded by
  MCIN/AEI/10.13039/501100011033 as well as Junta de Andaluc{\'i}a
  (grant FQM-225).
\end{acknowledgments}

\appendix

\section{General Lorentz structure \label{app:decomp}}

By Lorentz covariance, the matrix element (we skip the isospin indices for the simplicity of notation)
\begin{eqnarray}
M^\mu = \langle\pi(p+q)|\overline{\psi}(-\tfrac{\lambda}{2}n)\, \gamma^\mu  \psi(\tfrac{\lambda}{2}n)|\pi(p)\rangle 
\end{eqnarray}
can be decomposed as
\begin{eqnarray}
M^\mu =A p^\mu +  B q^\mu +C n^\mu,
\end{eqnarray}
thus leading to three independent amplitudes, $A$, $B$, and $C$. Explicitly,
\begin{eqnarray}
&&A=\frac{4  \left(\zeta ^2 M_p-n^2 t\right)-2 M_q \left(n^2 t-2 \zeta \right)-2 (\zeta -2) t M_n}{4 m_{\pi }^2 \left(\zeta^2-n^2 t\right)+t \left(-4 \zeta +n^2 t+4\right)}, \nonumber \\
&& B=\frac{ \left(4 \zeta\! -\!2 n^2 t\right)\!M_p\!-\!4 \left(m_{\pi }^2 n^2\!-\!1\right) \!M_q\!+\!2  \left(t\!-\!2 \zeta  m_{\pi}^2\right)\!M_n}{4m_{\pi }^2 \left(\zeta ^2-n^2 t\right)+t \left(-4 \zeta +n^2 t+4\right)}, \nonumber \\
&& C = \frac{2 (2-\zeta ) t M_p+2  \left(t-2 \zeta  m_{\pi}^2\right)M_q+t \left(t-4 m_{\pi }^2\right) M_n}{4m_{\pi }^2 \left(\zeta ^2-n^2 t\right)+t \left(-4 \zeta +n^2 t+4\right)}, \nonumber \\
\end{eqnarray}
where $M_a=M^\mu a_\mu$
Our choice of Eqs.~(\ref{eq:Hdef},\ref{eq:Edef}) corresponds, along the lines of the original proposal by Ji~\cite{Ji:2013dva}, to the combination 
\begin{eqnarray}
n_\mu M^\mu=A-\zeta B +n^2 C. \label{eq:jidef}
\end{eqnarray}
However, as suggested in~\cite{Orginos:2017kos}, in view of the lattice QCD implementations it may be advantageous to project out the $C$ term, which would lead to 
\begin{eqnarray}
&& A-\zeta B = \nonumber \\ 
&& \frac{2 n^2 \left[(\zeta \!-\!2) t M_p-\left(t\!-\!2 \zeta  m_{\pi }^2\right)M_q \right]+4 M_n \left(\zeta ^2 m_{\pi }^2\!-\!\zeta  t\!+\!t\right)}{4 m_{\pi }^2 \left(\zeta ^2-n^2 t\right)+t \left(-4 \zeta +n^2 t+4\right)}. \nonumber \\
\label{eq:proj}
\end{eqnarray}
Also, the $C$ term of the amplitude is subleading in the twist expansion~\cite{Orginos:2017kos}. 

Although a consideration of the general or of the projected case~(\ref{eq:proj}) is certainly possible, we do not pursue it here due to algebraic complications, and hold on to the definition~(\ref{eq:jidef}).

\section{Kinematics \label{app:kin}}

Whereas our calculations are fully covariant, it is worthwhile to consider specific assignments for the momenta, having in mind possible lattice implementations where a particular reference frame must be specified. Without a loss of generality, we may pick up a frame where the three Lorenz vectors have the $t,x,y,z$ coordinates taken as
\begin{eqnarray}
n&=&(n_0, 0, 0, n_3), \nonumber \\
p&=&(p_0, 0, 0, p_3), \nonumber \\
q&=&(q_0, q_T, 0, q_3). \label{eq:genmom}
\end{eqnarray}
For the $n$ vector, we use the condition (\ref{eq:nPz}), where $P_z$ is a {\em parameter} (not necessarily equal to $p_3$) controlling how far $n$ is from the null vector case $n^2=0$, which corresponds to the limit of  $P_z\to\infty$ (the GPD case).
With the choice~(\ref{eq:genmom}) we have 
\begin{eqnarray}
\epsilon^{n p q \nu} = \delta_{\nu 2} ( p_0 n_3 - p_3 n_0) q_T. 
\end{eqnarray}

Treating $p_0$ and $p_3$ as known variables, with \mbox{$p_0^2=m_\pi^2+p_3^2$}, and using the total of five conditions from Eq.~(\ref{eq:momcond}) and (\ref{eq:nPz}), we solve the system for $n_0$, $n_3$, $q_0$, $q_T$, and $q_3$, with the result 
\begin{eqnarray}
n_0&=&\frac{p_0}{m_\pi^2}-\frac{p_3 \sqrt{m_\pi^2+P_z^2}}{m_\pi^2 P_z}, \nonumber \\
n_3&=&\frac{p_3}{m_\pi^2}-\frac{p_0 \sqrt{m_\pi^2+P_z^2}}{m_\pi^2 P_z}, \nonumber \\
q_0&=&-\frac{p_0 t}{2 m_\pi^2}+\frac{p_3 P_z \left(t-2 \zeta  m_\pi^2\right)}{2 m_\pi^2\sqrt{m_\pi^2+P_z^2}}, \nonumber \\
q_3&=&-\frac{p_3 t}{2 m_\pi^2}+\frac{p_0 P_z \left(t-2 \zeta  m_\pi^2\right)}{2 m_\pi^2\sqrt{m_\pi^2+P_z^2}}, \nonumber \\
q_T^2&=&\frac{t \left[ \tfrac{1}{4}t- (1-\zeta ) P_z^2\right]-m_\pi^2 \left(\zeta ^2 P_z^2+t\right)}{m_\pi^2+P_z^2}. \label{eq:kgen}
\end{eqnarray}

For the specific choice $P_z=p_3$ \cite{Ji:2013dva} Eqs.~(\ref{eq:kgen}) reduce to
\begin{eqnarray}
n_0&=&0, \;\;\; n_3=-\frac{1}{P_z}, \nonumber \\
q_0&=&-\frac{2 \zeta  P_z^2+t}{2 p_0}, \;\;\; q_3=-\zeta P_z, \nonumber \\
q_T^2&=&\frac{\left(2 \zeta  P_z^2+t\right)^2}{4 p_0^2}-\zeta ^2 P_z^2-t, \label{eq:kji}
\end{eqnarray}
where $n$ is aligned with the $z$ direction.
Another potentially useful case is for the initial pion at rest, when
\begin{eqnarray}
n_0&=&\frac{1}{m_\pi}, \;\;\; n_3=-\frac{\sqrt{m_\pi^2+P_z^2}}{m_\pi P_z}, \nonumber \\
q_0&=&-\frac{t}{2 m_\pi}, \;\;\; q_3=\frac{P_z \left(t-2 \zeta  m_\pi^2\right)}{2 m_\pi \sqrt{m_\pi^2+P_z^2}}, \nonumber \\
q_T^2&=&\frac{t \left(\tfrac{1}{4}t-(1-\zeta) P_z^2\right)- m_\pi^2 \left(\zeta ^2 P_z^2+t\right)}{m_\pi^2+P_z^2}. \label{eq:krest}
\end{eqnarray}

The case of GPDs corresponds to the null vector $n$, which can be achieved from Eqs.~(\ref{eq:kgen}) by taking the limit $P_z\to\infty$, with the result
\begin{eqnarray}
n_0&=&\frac{1}{p_0+p_3}, \;\;\; n_3=-\frac{1}{p_0+p_3}, \nonumber \\
q_0&=&-\frac{t}{2 (p_0+p_3)}-\zeta  p_3, \;\;\; q_3=\frac{t}{2 (p_0+p_3)}-\zeta  p_0, \nonumber \\
q_T^2&=&(\zeta -1) t-\zeta ^2 m_{\pi }^2. \label{eq:kgpd}
\end{eqnarray}

For a physical process all momenta must have real coordinates, in particular $q_T$ must be real. This leads to constraints, which for the qGPD case at various values of  $P_z$ take the form
\begin{eqnarray}
&&\frac{t}{2}  \le  m_\pi^2+(1-\zeta) P_z^2 -\sqrt{\left(m_\pi^2+P_z^2\right) \left(m_\pi^2+(1-\zeta )^2 P_z^2\right)},\nonumber \\
&& \hspace{0.7cm} {\rm or} \label{eq:condg} \\
&&\frac{t}{2}  \ge  m_\pi^2+(1-\zeta) P_z^2 +\sqrt{\left(m_\pi^2+P_z^2\right) \left(m_\pi^2+(1-\zeta )^2 P_z^2\right)}. \nonumber
\end{eqnarray}
For the special case of GPD ($P_z\to\infty$) conditions~(\ref{eq:condg}) reduce to 
\begin{eqnarray}
t  \le  - \frac{m_\pi^2 \zeta^2}{1-\zeta}. \label{eq:condGPD} 
\end{eqnarray}
This shows that the maximum value of $t$ is (for $\zeta>0$) strictly less than 0. The reason is that the momentum transfer along $n$ (for $\zeta>0$) brings in a negative contribution to $t$. 

\section{Evaluation of the loop integrals \label{app:loops}}

In this Appendix we explain for completeness the evaluation of the basic loop integrals. The procedure, using standard methods, follows closely Ref.~\cite{Broniowski:2007si}. We encounter two types of scalar loop integrals, the two-point function $I$ and the three-point function $J$, defined below. They are evaluated in the Euclidean space using the Schwinger parameterization of the scalar propagators,
\begin{eqnarray}
S_k=\frac{1}{D_k}=\int_0^\infty e^{-a (k^2+M^2)}. \label{eq:sch}
\end{eqnarray}
Note that in the Euclidean notation used in the Appendices,  $p^2=(p+q)^2=-m_\pi^2$, $q^2=-t$, and $n^2 \ge 0$.

\subsection{Two-point function \label{app:2p}}

 With the representation~(\ref{eq:sch}) we have
\begin{widetext}
\begin{eqnarray}
    I(y,\kappa,l^2,n^2) &=& 4 N_c g_{\pi q}^2 \int\frac{d^4k}{(2\pi)^4} \frac{\delta(k\cdot n-y)}{D_{k}D_{k-l}} \label{eq:iii} \\
    &=& 4 N_c g_{\pi q}^2 \int\frac{d^4k}{(2\pi)^4} \int\frac{d\lambda}{2\pi}e^{i\lambda(k\cdot n-y)}\int_0^\infty \!\! d\alpha \int_0^\infty \!\! d\beta \, e^{-\alpha(k^2+M^2)-\beta((k-l)^2+M^2)}\nonumber\\
    &=& 4 N_c g_{\pi q}^2 \int\frac{d^4k}{(2\pi)^4} \int\frac{d\lambda}{2\pi}\int_0^\infty  \!\! d\alpha \int_0^\infty \!\! d\beta \exp\Big[-(\alpha+\beta)(k^{\prime 2}+M^2)-\frac{\lambda^2 n^2}{4(\alpha+\beta)}+i\frac{\lambda\beta \kappa}{\alpha+\beta}-\frac{\alpha \beta l^2}{\alpha+\beta}-i\lambda y\Big]. \nonumber
\end{eqnarray}
\end{widetext}
where the shifted momentum is $k^\prime=k-i\frac{\lambda n}{2(\alpha+\beta)}-\frac{\beta l}{\alpha+\beta}$. 

First, we notice that in the case of $n^2=0$, the $\lambda$ integration yields $\delta(y-\frac{\kappa\beta}{\alpha+\beta})$, 
hence the proper support $\theta[x(\kappa-x)]$ for the momentum fraction $x=y$ follows~\cite{Broniowski:2007si}. However, when (Euclidean) $n^2 > 0$, the $\lambda$ integral is over a Gaussian with a spread proportional to $\frac{1}{n^2}$. Thus, after the $\lambda$ integration, we obtain a Gaussian function in $y$ whose width is proportional to $n^2$. Clearly, in the limit of $n^2\to 0$ we retrieve the result given in Ref.~\cite{Broniowski:2007si}.

Next, we use the following change of variables:
\begin{eqnarray}
&& k_T^2= k_1^{\prime 2} + k_2^{\prime 2}, \;\;\; K^2= k_0^{\prime 2} + k_3^{\prime 2}, \\
&& s=\alpha+\beta, \;\;\; \psi=\frac{\beta}{s},  \nonumber 
\end{eqnarray}
with $dk_1 dk_2= \pi dk_T^2$, $dk_0 dk_3= \pi dK^2$, and  $d\alpha\, d\beta=s \,ds \,d\psi$. The integration over $K^2$ and $s$ yields
\begin{eqnarray}
&& I(y,\kappa,l^2,n^2) = \frac{N_c g_{\pi q}^2}{8\pi^2 \sqrt{n^2}} \times \label{eq:ii} \\
&& \int_0^\infty \!\!\! dk_T^2\int_0^1 \!\! d\psi \frac{1}{\big[k_T^2 +M^2+\psi(1-\psi)l^2+\frac{1}{n^2}(y-\psi\kappa)^2 \big]^{3/2}}, \nonumber
\end{eqnarray}
while the further integration over $\psi$ gives the result
\begin{eqnarray}
&& I(y,\kappa,l^2,n^2) = \frac{N_c\, g_{\pi q}^2}{4\pi^2 f^2} \int_0^\infty \!\! dk_T^2 \times \label{eq:igen}\\
&& \frac{\frac{2 \kappa  y- l^2 n^2}{\sqrt{y^2+n^2(k_T^2+w^2)}}+\frac{2 \kappa  (\kappa -y)-l^2 n^2}{\sqrt{(\kappa -y)^2+n^2(k_T^2+w^2)}}}
{4 \left[\kappa ^2 (k_T^2+w^2)+l^2 y (\kappa -y) \right]-l^2 n^2 \left[ l^2+4 \left(k_T^2+w^2\right)\right]} \nonumber
\end{eqnarray}
Note the desired symmetry $y \leftrightarrow \kappa-y$.

In the limit of $n^2=0$ and $y=x$ we promptly recover the result of Ref.~\cite{Broniowski:2007si}:
\begin{eqnarray}
 I(y,\kappa,l^2,n^2=0) &=& \frac{N_c\, g_{\pi q}^2 \theta[x(\kappa-x)]}{4\pi^2  |\kappa |} \times \label{eq:ilim} \\
&& \int_0^\infty \!\!\! dk_T^2 \frac{1}{k_T^2+M^2 +\frac{x}{\kappa} \left(1- \frac{x}{\kappa} \right)l^2} \nonumber .
\end{eqnarray}
Since the remaining integration over $k_T$ in Eq.~(\ref{eq:ilim}) is logarithmically divergent, it can only be carried out after a suitable regularization. 

However, curiously, the $k_T$ integration can be carried out in Eq.~(\ref{eq:igen}) where a nonzero $n^2$ acts as a regulator, and asymptotically the integrand in Eq.~(\ref{eq:igen}) behaves as $1/k_T^3$. The result of the integration is (we show it explicitly for the case of $l^2=-m_\pi^2$ encountered in our analysis)
\begin{eqnarray}
 &&I(y,\kappa,-m_\pi^2,n^2)=\frac{N_c g_{\pi q}^2}{4 \pi ^2} \times \label{eq:fa}\\
 &&\frac{\log \left[ \frac{2\sqrt{\left(\kappa ^2+m_\pi^2 n^2\right) \left(n^2 w^2+(y-\kappa )^2\right)}-2\kappa (y-\kappa) +m_\pi^2 n^2}{2 \sqrt{\left(\kappa ^2+m_\pi^2 n^2\right) \left(n^2 w^2+y^2\right)}-2 \kappa  y -m_\pi^2 n^2}\right]}{\sqrt{\kappa^2+m_\pi^2 n^2}}. \nonumber
\end{eqnarray}
This expression can be shown to be symmetric with respect to the replacement $y \leftrightarrow \kappa-y$. It exhibits the expected quark-antiquark production cut for $m_\pi > 2M$. The asymptotic behavior at large $|y|$ is $1/|y|$.

Since for $n^2>0$ the $k_T$ integration can be carried out, we can rewrite Eq.~(\ref{eq:ii}) as
\begin{eqnarray}
&& I(y,\kappa,l^2,n^2) = \frac{N_c g_{\pi q}^2}{4\pi^2 \sqrt{n^2}} \times \label{eq:ii0} \\
&& \int_0^1 \!\! d\psi \frac{1}{\big[M^2+\psi(1-\psi)l^2+\frac{1}{n^2}(y-\psi\kappa)^2 \big]^{1/2}}. \nonumber
\end{eqnarray}

Coming back to Eq.~(\ref{eq:ilim}), it can be promptly derived from Eq.~(\ref{eq:ii}) by noticing that it contains the distribution
\begin{eqnarray}
\lim_{n^2 \to 0} \frac{1}{\sqrt{n^2} \left (A^2 + \frac{B^2}{n^2} \right)^{3/2}}=\frac{2}{A^2}\delta(B). \label{eq:distr}
\end{eqnarray}

\subsection{Three-point function}

The scalar triangle integral is defined as
\begin{eqnarray}
&& J(y,\kappa,\kappa^\prime,l^2,l^{\prime 2},l\cdot l^\prime, n^2)= \\
&& \hspace{1.5cm} 4 N_c g_{\pi q}^2 \int \frac{d^4k}{(2\pi)^2} \frac{\delta(k\cdot n-y)}{D_k D_{k-l}D_{k-l^\prime}}. \nonumber 
\end{eqnarray}
Following the procedure from Appendix~\ref{app:2p} for 
\begin{eqnarray}
\hspace{-10mm} J&=&4 N_c g_{\pi q}^2\int\frac{d^4k}{(2\pi)^4} \int\frac{d\lambda}{2\pi}\int_0^\infty  \!\! d\alpha \int_0^\infty \!\! d\beta \int_0^\infty \!\! d\gamma \times \nonumber \\
\hspace{-10mm}&& e^{i\lambda[k\cdot n -y]-\alpha[k^2+M^2]-\beta[(k-l)^2+M^2]-\gamma[(k-l')^2+M^2]} 
\end{eqnarray}
with
\begin{eqnarray}
&&s=\alpha+\beta+\gamma, \;\; \psi=\beta/s, \;\; \tau=\gamma/s, \\
&& d\alpha\, d\beta\, d\gamma=s^2 ds \, d\psi \, d\tau, \nonumber
\end{eqnarray}
we find
\begin{widetext}
\begin{eqnarray}
    J &=& \frac{3N_c g_{\pi q}^2}{16\pi^2 \sqrt{n^2}}\int_0^1d\psi\int_0^1d\tau \int_0^\infty \!\! dk_T^2 \frac{\theta(1-\psi-\tau)}{\left[k_T^2 +M^2+ \psi(1-\psi)l^2+\tau(1-\tau)l'^2 -2 \psi\tau l\cdot l' +\frac{1}{n^2}(y-\kappa \psi-\kappa' \tau)^2\right]^{5/2}} \nonumber \\
    &=&  \frac{N_c g_{\pi q}^2}{8\pi^2 \sqrt{n^2}}\int_0^1d\psi\int_0^1d\tau \frac{\theta(1-\psi-\tau)}{\left[M^2+ \psi(1-\psi)l^2+\tau(1-\tau)l'^2 -2 \psi\tau l\cdot l' +\frac{1}{n^2}(y-\kappa \psi-\kappa' \tau)^2\right]^{3/2}}. \label{eq:jj}
\end{eqnarray}
\end{widetext}
The integral over $k_T^2$ is finite, hence above it could have been carried out. The integrations over $\psi$ and $\tau$ are analytic, but the final results are very lengthy and not instructive, so we do not quote them. 

Using Eq.~(\ref{eq:distr}) we find that in the limit of $n^2\to 0$ 
\begin{eqnarray}
&&\hspace{-11mm} J(x,\kappa,\kappa^\prime,l^2,l^{\prime 2},l\cdot l^\prime, 0)= 
    \frac{N_c g_{\pi q}^2}{4\pi^2}\int_0^1d\psi\int_0^1d\tau \times \nonumber \\
&&\hspace{-3mm}\frac{\theta(1-\psi-\tau) \delta(x-\kappa \psi-\kappa' \tau)}{M^2+ \psi(1-\psi)l^2+\tau(1-\tau)l'^2 -2 \psi\tau l\cdot l'},
\end{eqnarray}
which agrees with the result of Ref.~\cite{Broniowski:2007si}.

\section{GPDs for $m_\pi=0$ and $t=0$ \label{app:lim}}

The GPD case ($n^2=0$) for $m_\pi=0$ and $t=0$ yields very simple expressions. For the GPDs we have~\cite{Broniowski:2007si}
\begin{eqnarray}
H^{I=1}(X,\xi)&=& \theta(1-X^2), \label{eq:h1s}  \\
H^{I=0}(X,\xi)&=& {\rm sgn}(X)\theta(1-X^2)\theta(X^2 - \xi^2), \nonumber
\end{eqnarray}
whereas for the tGPDs~\cite{Dorokhov:2011ew}
\begin{eqnarray}
&& E^{I=1}(X,\xi)= {\cal N}\,\theta(1-X^2) \label{eq:e1s}\\
&&               ~~~~~\times  \left [ \theta (| X| -\xi )\frac{| X| -1}{\xi -1}+\theta \left(\xi^2-X^2\right) \right ], \nonumber \\
&& E^{I=0}(X,\xi)= {\cal N}\theta(1-X^2) \nonumber \\
&&               ~~~~~\times \left [  {\rm sgn}(X) \theta (| X| -\xi )\frac{(| X| -1) }{\xi-1}+\theta \left(\xi ^2-X^2\right)\frac{X }{\xi } \right ]. \nonumber 
\end{eqnarray}
The normalization constant is 
\begin{eqnarray}
{\cal N}=\frac{N_c g_{\pi q}^2 M}{4\pi^2} \left . \frac{1}{M^2} \right |_{\rm reg.}, \label{eq:normN}
\end{eqnarray}
where in the adopted PV regularization~(\ref{eq:pvr})
\begin{eqnarray}
 \left . \frac{1}{M^2} \right |_{\rm reg.}=\frac{\Lambda^4}{M^2 (\Lambda^2+M^2)^2}.
\end{eqnarray}
The corresponding reduced ITDs of Eq.~(\ref{eq:redI}) are given in Eq.~(\ref{eq:iofgpd}).

\section{Polynomiality \label{app:poly}}

The moments or $I$ and $J$ functions with respect to $y$ are defined as 
\begin{eqnarray}
\langle y^m I \rangle &=& \int_{-\infty}^\infty dy \, y^m  I(y,\kappa,l^2,n^2), \\
\langle y^m J \rangle &=& \int_{-\infty}^\infty dy \, y^m J(y,\kappa,\kappa^\prime,l^2,l^{\prime 2},l\cdot l^\prime, n^2), \nonumber
\end{eqnarray}
with $m=0,1,2,\dots$.
Taking Eq.~(\ref{eq:ii}) and changing the integration variable to $y'=(y-\kappa \psi)/\sqrt{n^2}$ we can write
\begin{eqnarray}
&& \langle y^m I \rangle = \frac{N_c g_{\pi q}^2}{8\pi^2} \times \label{eq:iiy} \\
&& \int_{-\infty}^\infty dy'  \int_0^\infty \!\!\! dk_T^2\int_0^1 \!\! d\psi \frac{(\sqrt{n^2} y'+\kappa \psi)^m}{\big[k_T^2 +M^2+\psi(1-\psi)l^2+y'^2 \big]^{3/2}}.\nonumber
\end{eqnarray} 
If the integral over $y'$ exists, it is a polynomial of degree $m$ in $\kappa$ with coefficients given by functions of $l^2$ and $n^2$. 

Similarly, with Eq.~(\ref{eq:jj}) and the change of variables $y'=(y-\kappa \psi-\kappa' \tau)/\sqrt{n^2}$, we find that $\langle y^m J \rangle$ is a polynomial 
in variables $\kappa$ and $\kappa'$ of degree $m$ with coefficients given by functions of $l^2$, $l'^2$, $l\cdot l'$, and $n^2$.

After same work, polynomiality of the basic loop integrals translates into polynomiality of the $y$-moments of qGPDs and qtGPDs for the on-shell pion, which (if exist) are polynomials in $\xi$ with coefficients given by the generalized quasi form factors, which are functions of $t$ and, in general, $n^2$. 

\section{Radyushkin's relations for the scalar one-loop functions \label{app:rad}}

The relations derived by Radyushkin~\cite{Radyushkin:2016hsy,Radyushkin:2017cyf,Radyushkin:2017lvu,Radyushkin:2017gjd}, following entirely from the Lorentz covariance, link nontrivially the quasi-distributions $\tilde{q}(y,n^2)$ to the $k_T$-unintegrated distributions $q(x,k_T^2)$, namely
\begin{eqnarray}
\hspace{-7mm} \tilde{q}(y,n^2)= \frac{1}{\sqrt{n^2}} \int dk_1 \int dx \, q[x,k_1^2-\frac{(x-y)^2}{n^2}].
\end{eqnarray} 
Here we verify explicitly that these relations hold separately for the basic $n$-point functions, and hence generalize for the considered quasi-(t)GPDs.

Taking Eq.~(\ref{eq:iii}) and integrating it over $dK^2$ we can write
\begin{eqnarray}
    && I(y,\kappa,l^2,n^2) = \frac{N_c g_{\pi q}^2}{4\pi^2} \int_0^\infty \!\!\! du \int \!\! \frac{d\lambda}{2\pi}\int_0^\infty \!\! \!ds \int_0^1 \!\! \! d\psi \times \label{eq:iR}  \\ 
    && \exp\Big[-s(u+M^2)-\frac{\lambda^2 n^2}{4s}+i \lambda\psi \kappa-s \psi(1-\psi) l^2 -i\lambda y\Big], \nonumber
\end{eqnarray}
where $u=k_T^2$. On the other hand, the $k_T$-unintegrated expression with $\bm{k_T}=(k_1,0)$ and $n^2=0$ reads
\begin{eqnarray}
    && I(x,\kappa,l^2,k_1^2) = \frac{N_c g_{\pi q}^2}{4\pi^3} \int_ 0^\infty dU \int \!\! \frac{d\lambda}{2\pi}\int_0^\infty \!\!\! s\,ds \int_0^1 \!\! \! d\psi \times  \nonumber \\ 
    && \exp\Big[-s(U+k_1^2+M^2)+i \lambda\psi \kappa-s \psi(1-\psi) l^2 -i\lambda x\Big], \nonumber\\
    \label{eq:iR2} 
\end{eqnarray}
where $U=K^2$. Using the identity
\begin{eqnarray}
\frac{1}{\sqrt{n^2}} \int dk_1 {e^{-s [k_1^2+\frac{(x-y)^2}{n^2}]-i \lambda  x}}=\frac{\pi}{s}  e^{-\frac{\lambda^2}{4s} - i \lambda y}
\end{eqnarray}
we immediately verify that 
\begin{eqnarray}
 I(y,\kappa,l^2,n^2)= \frac{1}{\sqrt{n^2}} \int \!\! dk_1 \!\! \int \!\! dx \,  I[ x,\kappa,l^2,k_1^2+P_z^2(x-y)^2]. \nonumber \\
\end{eqnarray}

A completely analogous derivation holds for the three-point function $J$, as well as for any scalar one-loop $n$-point function.

\section{One-loop functions for the generalized Ioffe-time distributions and generalized pseudo-distributions \label{app:ioffe}}

The Ioffe-time representation of the scalar one-loop function is obtained from Eq.~(\ref{eq:ii0}) as the Fourier transform
\begin{eqnarray}
&& I^I(-\nu,\kappa,l^2,-z^2)=\int dy\, e^{i \nu y} I(y,\kappa,l^2,-z^2/\nu^2)   \\
&&\hspace{0.1cm}  =\frac{N_c g_{\pi q}^2}{2\pi^2} \int_0^1 \!\! d\psi \, e^{i\nu\kappa\psi}  K_0 \left( \sqrt{-z^2 [M^2+\psi(1-\psi)l^2]} \right). \nonumber
\end{eqnarray}
In the evaluation we have used the covariant formula $n^2=-z^2/\nu^2$, 
The corresponding pseudo-distribution is
\begin{eqnarray}
&& I^P(x,\kappa,l^2,-z^2)=\int \frac{d\nu}{2\pi}\, e^{-i \nu x} I^I(\nu,\kappa,l^2,-z^2)  =\frac{N_c g_{\pi q}^2}{2\pi^2} \times \nonumber  \\
&&\hspace{5mm}   \int \frac{d\nu}{2\pi} \int_0^1 \!\! d\psi \, e^{-i\nu(x-\kappa\psi})   K_0 \left( \sqrt{-z^2 [M^2+\psi(1-\psi)l^2]} \right) \nonumber \\
&&\hspace{5mm}  =\frac{N_c g_{\pi q}^2\theta[x(\kappa-x)]}{2\pi^2 | \kappa |} K_0 \left( \sqrt{-z^2 [M^2+\tfrac{x}{\kappa}(1-\tfrac{x}{\kappa})l^2]} \right), \nonumber \\ \label{eq:psI}
\end{eqnarray}
where in the last line we have applied the $\delta(x-\kappa \psi)$ function appearing in the second line. The symbol $K_l$ denotes the modified Bessel function of the second kind.

Similarly, for the three-point function we find with Eq.~(\ref{eq:jj})
\begin{eqnarray}
&& J^I(-\nu,\kappa,\kappa^\prime,l^2,l^{\prime 2},l\cdot l^\prime, -z^2) =  \frac{N_c g_{\pi q}^2}{4\pi^2}\int_0^1\!\! d\psi\int_0^1\!\! d\tau \times \nonumber \\ 
&&\hspace{9mm} \theta(1-\psi-\tau)e^{i \nu (\kappa \psi+\kappa' \tau)} \frac{\sqrt{-z^2}}{A} K_1(\sqrt{-z^2 A}), \nonumber \\ \nonumber \\
&& A= M^2+ \psi(1-\psi)l^2+\tau(1-\tau)l'^2 -2 \psi\tau l\cdot l' 
\end{eqnarray}
and
\begin{eqnarray}
&& J^P(x,\kappa,\kappa^\prime,l^2,l^{\prime 2},l\cdot l^\prime, -z^2) =  \frac{N_c g_{\pi q}^2}{4\pi^2}\int_0^1\!\! d\psi\int_0^1\!\! d\tau \times \nonumber \\ 
&&\hspace{9mm} \theta(1-\psi-\tau)\delta (x-\kappa \psi-\kappa' \tau) \frac{\sqrt{-z^2}}{A} K_1(\sqrt{-z^2A}). \nonumber \\ \label{eq:psJ}
\end{eqnarray}
For our explicit kinematics, after doing the $\tau$ integration we get
\begin{eqnarray}
&& J^P(x,\zeta,1,t,m_\pi^2,-\tfrac{t}{2}, -z^2) =  \frac{N_c g_{\pi q}^2}{4\pi^2} \times \nonumber \\
&& \hspace{2mm}  \left [ \theta[x(\zeta-x)]\int_0^{\frac{x}{\zeta}}\!\! d\psi  + 
                                       \theta[(1-x)(x-\zeta)]\int_0^{\frac{1-x}{1-\zeta}}\!\! d\psi \right ] \times \nonumber \\ 
&&\hspace{9mm} \left . \frac{\sqrt{-z^2}}{A} K_1(\sqrt{-z^2A}) \right |_{\tau=x+\zeta \psi}, 
\end{eqnarray}
with
\begin{eqnarray}
&& A |_{\tau=x+\zeta \psi} = M^2 \\ 
&& \;\;\;\; - m_{\pi }^2 (1+\zeta  \psi -x) (x-\zeta  \psi)- t \psi  (1+\zeta  \psi -x-\psi). \nonumber \\ \nonumber
\end{eqnarray}

We also need (cf. Eq.~(\ref{tgpdeq})) the Fourier transforms of $J_1\equiv y\,J$, where
\begin{eqnarray}
&& \hspace{-9mm} J_1^I(-\nu,\kappa,\kappa^\prime,l^2,l^{\prime 2},l\cdot l^\prime, -z^2)) \nonumber \\ && =\int dy\, e^{i \nu y} y J =  
-i \frac{d}{d\nu} \int dy\,  e^{i \nu y} J \nonumber  \\
&& = -i \frac{d}{d\nu} J^I(-\nu,\kappa,\kappa^\prime,l^2,l^{\prime 2},l\cdot l^\prime, -z^2) ,
\end{eqnarray}
and therefore
\begin{eqnarray}
&& \hspace{-9mm} J_1^P(x,\kappa,\kappa^\prime,l^2,l^{\prime 2},l\cdot l^\prime, -z^2)) \nonumber \\
&& = x J^P(x,\kappa,\kappa^\prime,l^2,l^{\prime 2},l\cdot l^\prime, -z^2)).
\end{eqnarray}

\bibliography{Ref}

\end{document}